\documentclass[fleqn,usenatbib]{mnras}

\usepackage{newtxtext,newtxmath}

\usepackage[T1]{fontenc}
\usepackage{ae,aecompl}

\usepackage{graphicx}	
\usepackage{amsmath}	
\usepackage{amssymb}	

\newcommand{\fractal}{\%\mathrm{Frac}}

\newcommand{\dss}{$\delta$ Scuti stars}
\newcommand{\fractalpr}{\%\mathrm{Frac}_{\mathrm{pr}}}

\newcommand{\eqn} [1] {
\begin{equation}
#1
\end{equation}}

\newcommand{\eqna} [1] {
\begin{eqnarray}
#1
\end{eqnarray}}

\title[Fractal analysis of \dss]{Fractal analysis applied to light curves  of \dss \thanks{Based on data from COROT Archive at CAB.} }
\author[IAA/UGR Team]{
S. de Franciscis,$^{1}$\thanks{E-mail: sebas@iaa.es}
J. Pascual-Granado$^{1}$,
J.C. Su\'arez$^{2,1}$,
A. Garc\'ia Hern\'andez$^{2}$,
\newauthor{R. Garrido$^{1}$}
\\
$^{1}$Instituto de Astrof\'isica de Andaluc\'ia (IAA-CSIC). Glorieta de Astronom\'ia s$\setminus$n, 18008, Granada (Spain)\\
$^{2}$ Dept. F\'isica Te\'orica y del Cosmos. Univ. Granada (UGR). Av. de Fuente Nueva S$\setminus$N. 18071, Granada (Spain)}

\date{Accepted XXX. Received YYY; in original form ZZZ}

\pubyear{2015}

\begin{document}
\label{firstpage}
\pagerange{\pageref{firstpage}--\pageref{lastpage}}
\maketitle
%
%
\begin{abstract}
Fractal behaviour, i.e. scale invariance in spatio-temporal dynamics, have been found to describe and model many systems in nature, in particular fluid mechanics and geophysical related geometrical objects, like the convective boundary layer of cumulus cloud fields, topographic landscapes, solar granulation patterns, and observational astrophysical time series, like light curves of pulsating stars. 
%
%
%
%

The main interest in the study of fractal properties in such physical phenomena lies in the close relationships they have with chaotic and turbulent dynamic.

In this work we introduce some statistical tools for fractal analysis of light curves: Rescaled Range Analysis (R/S), Multifractal Spectra Analysis, and Coarse Graining Spectral Analysis (CGSA), an FFT based algorithm, which can discriminate in a time series the stochastic fractal power spectra from the harmonic one.

 An interesting application of fractal analysis in asteroseismology concerns the joint use of all these tools in order to develop classification criteria and algorithms for $\delta$-Scuti pulsating stars. In fact from the fractal and multi-fractal fingerprints in background noise of light curves we could infer on different mechanism of stellar dynamic, among them rotation,  modes excitation and magnetic activity.  
\end{abstract}

\begin{keywords}
stars: oscillations -- stars: variables: $\delta$-Scuti -- stars: activity
\end{keywords}


\section{Introduction}\label{sec:intro}
%
%
%
So far, fractal analysis has been applied with success to characterize light curves of solar-like stars \citep{NewSunsI,NewSunsIII} as well as M dwarfs \citep{NewSunsIV}. In \cite{NewSunsI, NewSunsIII} the authors characterized the solar-like stars most similar to the proper sun by studying their fractal and multifractal fingerprints, trying to finally link (multi)fractal parameters, rotational periods and flicker noise amplitudes. In this class of stars one of the possible sources of stochastic fluctuations has a simple interpretation; in fact here the mode excitation mechanism is called explicitly stochastic excitation mechanism, and it is modeled by a stochastically driven oscillator, whose driving force is a noise that models the dynamic of their typical convective thick outer layer pushing the inner radiative zone \citep{Gold77}. In \cite{NewSunsIV} the same authors studied fractal and multifractal properties of a M dwarfs dataset, finding interesting correlations with period of rotation and a previously well characterized magnetic activity index \citep{mathur2014photometric}.  M dwarfs are  low-mass stars, fully convective, and the observed magnetism is associated with the external convective envelope, where strong mass motions of conductive plasma induce a magnetic field through a cyclic dynamo process.   Also in this case there is a clear source of stochastic fluctuations: the signatures of both granulation and acoustic oscillations, excited in the outer convective zone, are included in the noise background of the time series at high frequencies, while the signatures at low frequencies can be caused by the rotational modulation of long-lived sunspots.

Inspired by these results, this work aims at detecting any fingerprint of fractal behavior in the light curves of \dss\ that may help to better understand their pulsational content, not yet well understood. Indeed, \dss\ might exhibit rather complex oscillation spectra due to the presence of a thin outer convective zone together with a rapid rotation.
These same characteristics made them especially interesting objects for testing the theories of angular momentum transport and chemicals \citep{Goupil05}. 
%

\begin{table*}
	\caption{Physical properties of the 6 \dss\  studied in this work ($7\cdot10^4-3.7\cdot10^5$ points and 32 seconds of time step). From left to right: ID represents the star identifier;  $\Delta\nu$ is the estimated large separation (in $\mu\mathrm{Hz}$); FT $\nu_{max}$ represents the Fourier frequency (in $\mu\mathrm{Hz}$) with the highest amplitude; $T_\mathrm{eff}$ is the effective temperature (in K); $\log g$  the surface gravity (in logarithmic scale);  $v \sin i$ is the projected equatorial velocity (in $\mathrm{km}\mathrm{s}^{-1}$); SpT represent the spectral type, and  $\tau$ is the observation time span in days. Uncertainties, when available, are indicated in parenthesis.}
    \label{tab:Stars}
	\begin{tabular}{lccccccc}
    \hline
ID & $\Delta\nu$ & FT $\nu_{max}$ & $T_\mathrm{eff}$ & $\log g$ & $v \sin i$ & SpT & $\tau$  \\
     & $\mu\mathrm{Hz}$&  $\mu\mathrm{Hz}$ & K & &  $\mathrm{km}\mathrm{s}^{-1}$ &  & days \\
\hline
HD174936 & 50 &	14,42 &	8000(200) &	4.08(0.2) &	169.7 &	A2 & 27.194 \\
HD174966 & 63 & 34 & 7555(50) &	4.21(0.05) & 126.1(1.2) & A3 & 27.197 \\
HD48784 & 54 & 14,73 & 6990(140) & 3.97 &108 & F0 & 25.305 \\ 
HD49434 & 23 & 11,41 & 7503(255) &	4.43(0.2)  &  85.7(4.3) & F1 V &	136.89 \\
HD50844	& 24 & 24,017 & 7400(200) & 3.6(0.2) & 58(2) & A2 &57.713 \\
HD50870	& 28 & 93,15 & 7660(250) & 3.68(0.25) &	37.5(2.5) &	A8 III & 114.413\\
\hline
\multicolumn{8}{l}{These data has been obtained from \citet{Javi18Prew} and references therein}\\
\end{tabular}
\end{table*}
The \dss\ are intermediate-mass ($1.5 - 3\,\mathrm{M}_\odot$), main-sequence A-F type pulsators, whose oscillations are maintained from the varying ionization of helium ($\kappa$ mechanism).
In the last years,  this  has been put into revision due to the detection of the so-called hybrid pulsators \citep[see e.g.][]{Ahmed2010, Uytterhoeven2011, Balona2015}, i.e. stars pulsating with both $\delta$ Scuti and $\gamma$ Doradus modes, whose oscillations are driven by convective blocking near the base of the convective envelopes resulting in the observed asymptotic gravity mode regime \citep{Guzik2000}.

In addition, ultra-precise data from CoRoT (Convection Rotation and planetary Transits, \citep{Baglin06}) and Kepler \citep{Gilli10} have revealed a flat plot of lower amplitude peaks  \citep[the so-called \emph{grass}, see e.g. ][]{Poretti09HD5084}, whose origin have been a matter of debate.  Among the candidates, it has been discussed a possible granulation background signal due to the effect of a thin outer convective layer \citep{KallingerMat10,Balona11}. Such a thin layer was claimed to be responsible for the excitation of solar-like oscillations in \dss\ \citep{Antoci11}. \footnote{This was not observed before the Kepler mission, so the presence of a convective envelope in the models has been questioned.} This competes with other possible effects like variations of the main peaks with time \citep{BarceloForteza15}, the existence of a magnetic field \citep{NeinerLampens15}, or the large number of chaotic modes \citep{LignieresGeor09} predicted in significantly deformed stars due to high rotation \citep[recently confirmed by][]{BarceloForteza17}.

All the aforementioned phenomena might be detected in some way by a fractal analysis to the light curve of \dss. Analyzing and understanding stochastic fluctuations and chaotic dynamic in the background spectra underlying the proper stellar pulsations modes could be very useful in the finding, modeling and description of the \dss\ mode distributions and amplitude modulation, and could give us good hints  on the proper excitation mechanism, as well as on magnetism and rotation.
The paper is organized as follows. First we specify the mathematical framework in which the fractal analysis is performed (Section~\ref{sec:safine}). Then, Section~\ref{sec:method} describes the observational data selected and and the methodology followed, including a brief description of all algorithms used for the fractal analysis. Discussion of the results are given in Section~\ref{sec:results}, and conclusions are summarized in Section~\ref{sec:conclusions}. Supplementary material is provided in appendices, including a short discussion of fractal analysis on simple theoretical models (Appendix~\ref{sec:Frac_Models}), the results obtained with multifractal analysis (Appendix~\ref{sec:MSS}), and a test of the sensitivity of the CGSA algorithm (Appendix~\ref{sec:CGSA_test}).

\section{Stellar light curves as self-affine time series}\label{sec:safine}

Fractals are mathematical sets defined \citep{manda77} through Self-Similarity (i.e. geometrical invariance under homogeneous scaling), emerging from infinite iteration rules, with no integer dimension that can be measured by the so called box counting method, generalizing the Euclidean dimension. Fractal dimension $D$ is the exponent of the power law dependence between the minimal number $N_r$ of boxes embedding the object and their linear dimension $r$, i.e.:
$$
D=\lim_{r\rightarrow 0}\frac{\log N_r}{\log \frac{1}{r}}.
$$
Fractal and multifractal behaviour have been found in several fluid mechanics dynamical systems: the convective boundary layer of cumulus cloud fields \citep{Pelletier97}, topographic landscapes and geological formation \citep{Scholz89,Meakin09}, rivers branching \citep{Tarboton88, DeBartolo00}, thin film growth by molecules deposition \citep{BarabasiBook95}. The main interest in the study of (multi)fractal properties in such physical phenomena lies in the close relationships they have with chaotic and turbulent dynamic \citep{Benzi84,Meneveau87,Dominguez92,Lyra98}.\\
Fractal spatio-temporal patterns and dynamics emerges in the context of critical phenomena, i.e. in the critical points of phase transitions (roughly speaking, the frontiers between two well characterized phases of a system). While in equilibrium statistical mechanics \citep{Chandler87} one needs an external fine tuning of relevant control parameters (like temperature or pressure) in order to put the system exactly in the critical point, there is another class of system, described by non-equilibrium statistical mechanics, where criticality is achieved spontaneously. In the former case we deal with Self Organized Criticality (SOC) \citep{Dickman00}, a kind of phenomenology vastly observed in natural systems, also in astrophysical context \citep{Charbo01,Chapman07,SOCAstro12}.
More generally, systems in a critical state are characterized by scale invariance, they have the same appearance at any spatial and temporal scale, and the characteristic functions of the system, as correlation functions or probability distributions, have a typical power law shape, i.e. the so called scale free distribution: $f(x)=x^{\zeta}$ \citep{Stanley72}.\\
On the other hand (multi)fractal behaviours could also emerge in low-dimensional dissipative systems at the onset of chaos \citep{Dominguez92,Lyra98}, including in strange non chaotic attractors, found in a natural system for the first time in some RR-Lyrae stars \citep{lindner15}.\\
In stellar physics fractal fingerprints in statistical observables (power law distributed) have been found in perimeter/area correlations \citep{SolGran86} and size and lifetime distributions of solar granules \citep{Lemme17}, in sunspot number and area variability \citep{zhou14, drozdz}, in magnetospheric substorms, auroras and flares \citep{SOCAstro12}, and finally in light curves from pulsating stars \citep{JaviFrac11, NewSunsI, NewSunsIII, NewSunsIV}.\\
The generalization of the concept of fractal from geometrical objects to time series, lead to the property of Self-Affinity:  a time series $y(t)$ is self-affine if it has the following inhomogeneous scaling relation \citep{Turco}:
\eqn{y(t)=\lambda^{H}y(\lambda t) \label{uno}}
where $\lambda$ is the scaling parameter and the constant H is the so-called Hausdorff exponent, characterizing long term correlations and the type of self-affinity in time series. Equation \ref{uno} has to be taken in statistical meaning, so that the scaling relationship holds when one performs appropriate measures on mean values over pairs of points at the same distance or over equal length subseries or windows etc.\\
The purpose of fractal analysis is to obtain statistical observables and to develop tools of measure in order to characterize the spatial and temporal evolution of correlations.\\
\section{Methodology}\label{sec:method}

In order to avoid spurious (instrumental) effects that may mimic the fractal behaviour in the stellar light curves we have selected a sample of 6 \dss\ observed by the CoRoT satellite, covering different observation times, and physical characteristics to ensure its representativity of the class (see details in Table~\ref{tab:Stars}). One possible source of false fractality might be the presence of gaps in the data, which occur at different scales and sizes. To overcome this problem gaps in the light curves (mostly instrumental) were treated with the gap-filling algorithm MIARMA \citep{Javi15miarma}, which maximizes the preservation of the frequency content. The light curves have a number of data points between $7.0\cdot10^{4}$ to $3.7\cdot10^{5}$, with 32 seconds of sampling. In addition, two of the stars, HD\,49434 and HD\,48784 exhibit mode frequencies in the  $\gamma$ Doradus regime \citep{ChapellierHD49434,BarceloForteza17}. These were included in the sample in order to compare the fractal behaviour of both $\delta$ Scuti and $\gamma$ Doradus within the current paradigm where hybrid pulsators are a challenging case of study (see Section~\ref{sec:intro}).
%
%
%
%
%
%
%
%
In the framework of self-affine time series, we worked out the light curve analysis with the statistical tools of Rescaled Range analysis (R/S) \citep{Turco, Kantel08} and Multifractal Singularity Spectra \cite{Kantel08}, already used in an astrophysical context \citep{JaviFrac11,NewSunsI,NewSunsIII}, and with Fourier power spectrum analysis, the most classical tool both in the stellar pulsation and in self-affine time series analysis. In addition we decided to explore, besides of Fourier amplitudes, the role played by the phases with the Coarse Grained Spectral Analysis (CGSA) \citep{Yama}, a novel tool which is able to quantify the contribution of stochastic fluctuations in time series. Multifractal Singularity Spectra (MSS) resulted to be very sensitive to the method, algorithm and parameters chosen, and it did not give sufficiently robust results, however we describe this tool and show some preliminary results in Appendix \ref{sec:MSS}.
\subsection{Rescaled Range Analysis}\label{sec:RRA}
An often used approach to the quantification of correlations in self-affine time series, determined by the H exponent in eq.\ref{uno} is Rescaled-Range analysis \citep{Kantel08}. Consider a discrete time series $y(i)$, ${i=1,2,3,\ldots N}$, and let split it into $N_l = N/l$ non-overlapping segments of size $l$. For each segment $y_n$ with $n=0,1,...N_l-1$, we consider its running sum time series relative to the mean value, i.e. $ys_{n}(j)=\sum_{i=1}^{j} \left(y_{n}(i)-\langle y_{n} \rangle \right)$ (where $j\in(1,\ldots,l)$ spans the whole segment). The Hurst exponent, $\alpha$, is obtained from
\begin{equation}\label{RS}
\lim_{l\rightarrow +\infty} \left( \frac{R_l}{S_l} \right)  = \lim_{l\rightarrow +\infty} RR(l) \propto l^{\alpha},
\end{equation}
here $R_{l}=\left< max(ys_{n})-min(ys_{n})\right>_{N_{l}}$, and $S_{l} = \left<\sigma_{n}\right>_{N_{l}}$ are the means over all segments of the difference between maxima and minima in the $n^{th}$ segment and its standard deviation, respectively. The exponent $\alpha$ is called Hurst exponent, from the name of its first developer of R/S technique.\\
%
%
It has been found empirically that many data sets in nature satisfy the power-law relation of eq.\ref{RS} , as river discharges, lake levels, tree ring thicknesses, sunspot numbers, and atmospheric temperature and pressure. R/S analysis is relevant in particular when one deals with quasi white fractional Gaussian noises (fGn), because here Hurst exponent is linearly related with power spectra exponent $\beta$, see next section \ref{sec:Spectra}.
%
%

\subsection{Spectral analysis }
\label{sec:Spectra}
Power spectrum, i.e the distribution of the signal power into frequency bins, of a self affine time series also has a power law behaviour:
\begin{equation}\label{PS}
S(\nu)\propto {\nu}^{-\beta}.
\end{equation}
%
%
%
%
From the relation $\beta\simeq2H+1$ we can obtain Hausdorff exponent, in the regime $\beta\in(1,3)$, i.e. for fractional Brownian motion, characterized by positive correlation and non-stationarity.\\
The exponent $\beta$ is also related to R/S coefficient $\alpha$ by $\beta\simeq 2\alpha-1$, when $\alpha$ is far away to the bounds of interval $\left[0,1\right]$, i.e. only for fractional Gaussian noises close to white Gaussian noise, characterized by stationarity with some slight positive ($\alpha>0.5$) or negative ($\alpha<0.5$) correlations \citep{Turco}. When, under the required conditions, the above relationship is not hold, it follows that the time series has multifractal properties \citep{Kantel08}. Thus, measuring both R/S $\alpha$ and power spectrum $\beta$ allow us first to discriminate if we are in the presence of fractional Brownian Motion or fractional Gaussian noise, secondly is a double check to characterize the amount of correlation in the case of Gaussian noises, and finally it could give us some hints on the multifractal nature of the series before a proper, and more complex, multifractal analysis is performed. For further details see appendix \ref{sec:Frac_Models}.
%
%
%
%
%
%

%
\begin{figure}
	\centering
		\includegraphics[width=9cm]{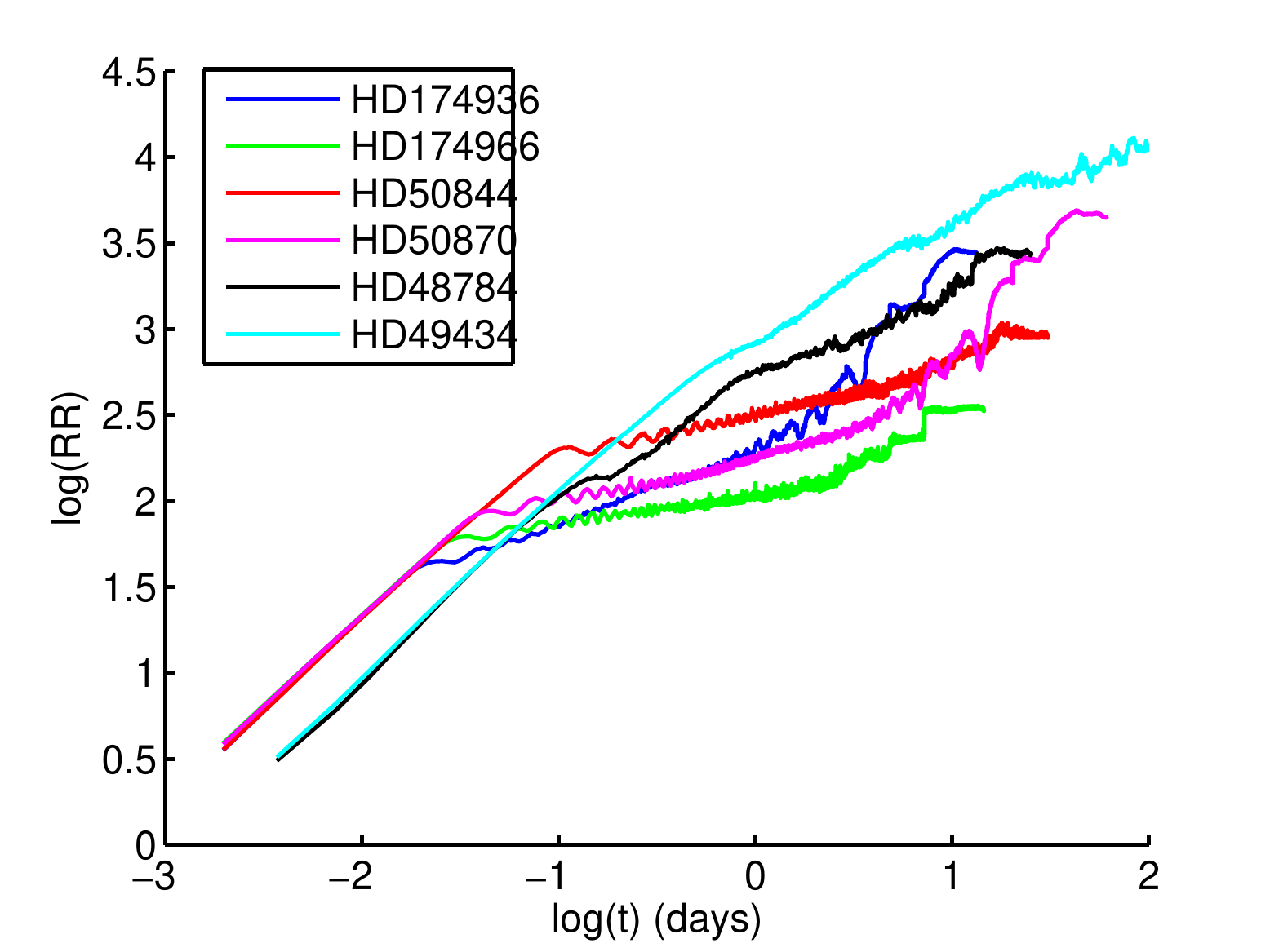}
	\caption{Rescaled Range analysis for the 6 CoRoT $\delta$-Scuti stars studied. Two different slopes are clearly visible indicating that two different regimes are present.}
	\label{RR_dScuti}
\end{figure}
\subsection{Coarse Graining Spectral Analysis}\label{sec:CGSA}

The Coarse Graining Spectral Analysis splits in a time series the self-affine component and the harmonic one, giving as output the percentage of (stochastic) fractal power in time series \citep{Yama}.  
While the majority of Fourier based analyses consider only the amplitudes of the harmonic components, disregarding half of the information resulting from a Fourier transform, i.e. the phases associated to each harmonic component, CGSA method focus also on the phase distribution.
CGSA is based on the consideration that in a self-similar time series the FFT phases follows a uniform distribution $\Theta_k\in \left[0, 2\pi \right]$.
We consider the original time series $y(i)$ %
%
%
%
%
and the series obtained by scaling $y(i)$ by a factor $2$ and $1/2$:
\eqna{
	y_2 &=& \{ y(2),y(4),y(6), \ldots\} \\
	y_{1/2}&=&\{y(1),y(1),y(2),y(2),\ldots \} . 
}
Next we cut those series in  $N_s$ partially overlapping subsets, each one having size the $90\%$ of the total length \footnote{$N_s$ is an additional free parameter.}. 
%
%
%
%
For each window $m$ we compute the auto-power spectrum $S_{yy,m}$ and the cross-power spectrum\footnote{Cross-power spectrum is defined as the Fourier transform of the cross-correlation function, i.e. $S_{XY}(k)=\sum_{n}\sum_{n'}X(n)Y(n+n')e^{ikn}$.} between the original series and the rescaled ones, i.e. $S_{yy_2,m}$ and $S_{yy_{\frac{1}{2}},m}$. 
If $y(i)$ is constituted by a sum of a few harmonics with fixed phase relationship it is possible to exploit the phase difference between windows $m-2$ and $m-1$ to orthogonalize $S_{yy_z,m}$ where $z \in \{1/2, 2\}$, with the rotating factor 
\eqn{S_{yy_z,m}^{o}(k)=S_{yy_z,m}(k)e^{-i\left[\pi/2-(\Theta_{m-1,k}-\Theta_{m-2,k})\right]}}
The residuals of such orthogonalization process are non zero in self-affine series, because any rescaled harmonic will find its counterpart in the original series and the phase relationships are always randomly distributed. Taking advantage of Schwartz's inequality we can calculate the fractal module cross correlations  
%
%
%
%
%
\eqna{
\langle || S_{yy_z,m-1}^{frac}(k) || \rangle_m &\equiv& \frac{\langle|| S_{yy_z,m-1}(k)\cdot S_{yy_z,m}^{o}(k) ||\rangle_m}{\langle S_{yy_z,m-1}(k)\rangle_m} \\
&\leq& \langle\mid\mid S_{yy_z,m}^{o}(k) \mid\mid \rangle_m.
}
Finally considering the possible distortions that could emerge by the finite size of the original series and the coarse graining of $y_{2}$ and $y_{1/2}$, we define the fractal power and the percentage of fractal power as
\eqna{
|| S^{frac}(k) || &\equiv& \sqrt{ || S_{yy_2}^{frac}(k) || \cdot || S_{yy_\frac{1}{2}}^{frac}(k) || }\\
\fractal &\equiv& \frac{\sum_{k}|| S^{frac}(k) ||}{\sum_{k}|| S(k) ||}
}

    

\begin{table*}
	\centering
	\caption{Coarse Grained Spectral Analysis $\fractal$, Rescaled Range 				exponent $\alpha$, Multifractal exponent $\alpha$ and Multifractal 				Singularity Spectra amplitude $\Delta\bar{\alpha}$  for 6 CoRoT 			\dss\ ($7\cdot10^4 - 3.7\cdot10^5$ points and 32 seconds of time sampling processed with 				MIARMA algorithm). $\fractal$ has been calculated for 				the original time series, and  a population of 10 phase 					randomized series ($\fractalpr$).  
    }    
    \label{tab:CGSA_6Stars}
	\begin{tabular}[!h]{lccccc}
\hline
ID& CGSA $\fractal$ & CGSA $\fractalpr$ & R/S\,$\alpha$ & MSS\,$\alpha$ & $\Delta \tilde{\alpha}$\\
\hline
HD174936 & 0.113 & 0.121$\pm$0.030 & 0.41 & 0.52 & 0.46 \\ 
HD174966 & 0.008 & 0.010$\pm$0.001 & 0.16 & 0.18 & 0.26 \\
HD48784 & 0.211 & 0.255$\pm$0.047 &  0.52 & 0.48 & 1.18 \\
HD49434 & 0.224 & 0.297$\pm$0.064 &  0.52 & 0.49 & 0.46 \\
HD50844 & 0.052 & 0.067$\pm$0.016 &  0.24 & 0.23 & 0.27 \\
HD50870 & 0.018 & 0.029$\pm$0.010 &  0.32 & 0.37 & 0.38 \\
\hline
\end{tabular}
\end{table*}
%


\begin{figure*}
	\centering
    \includegraphics[width=8cm]{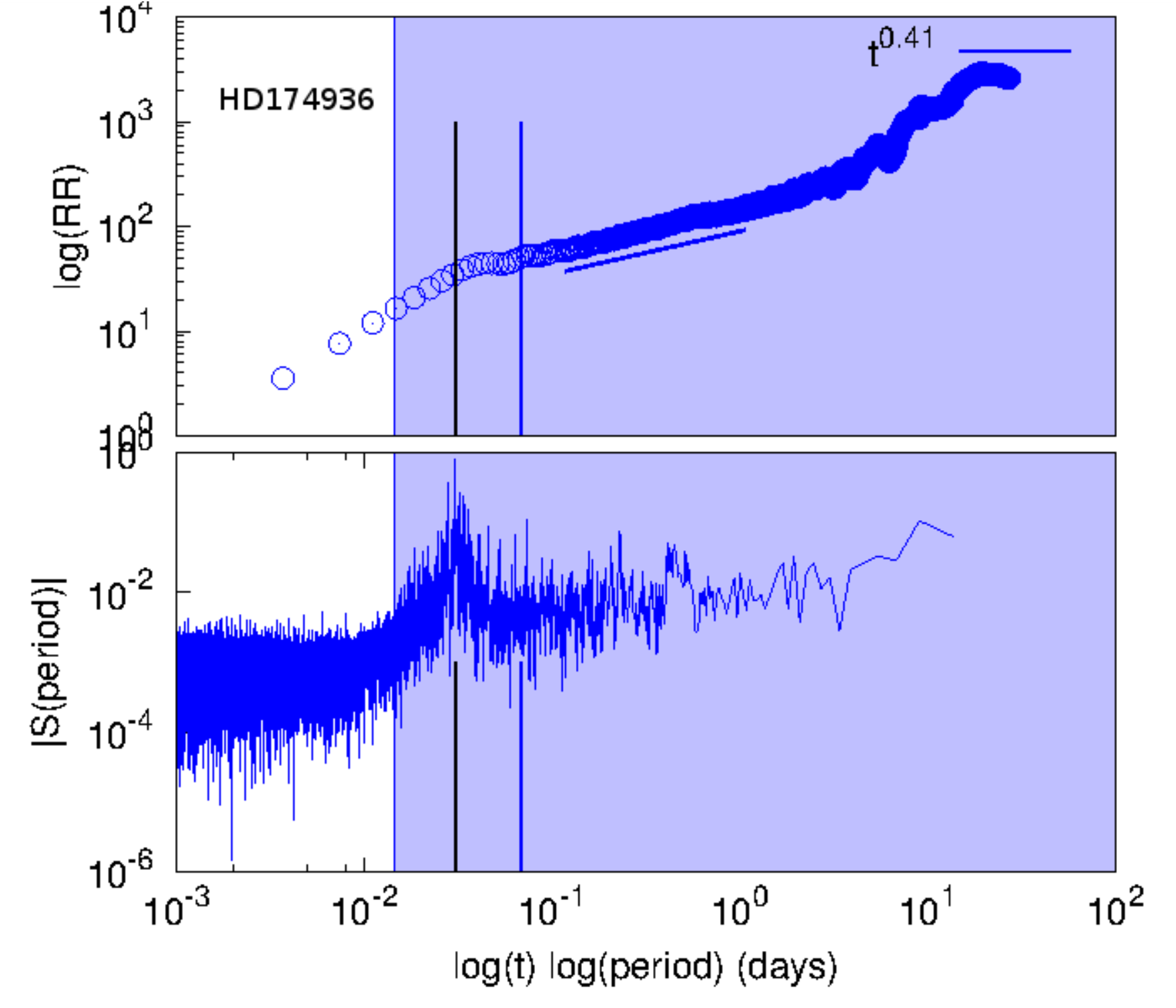}
    \includegraphics[width=8cm]{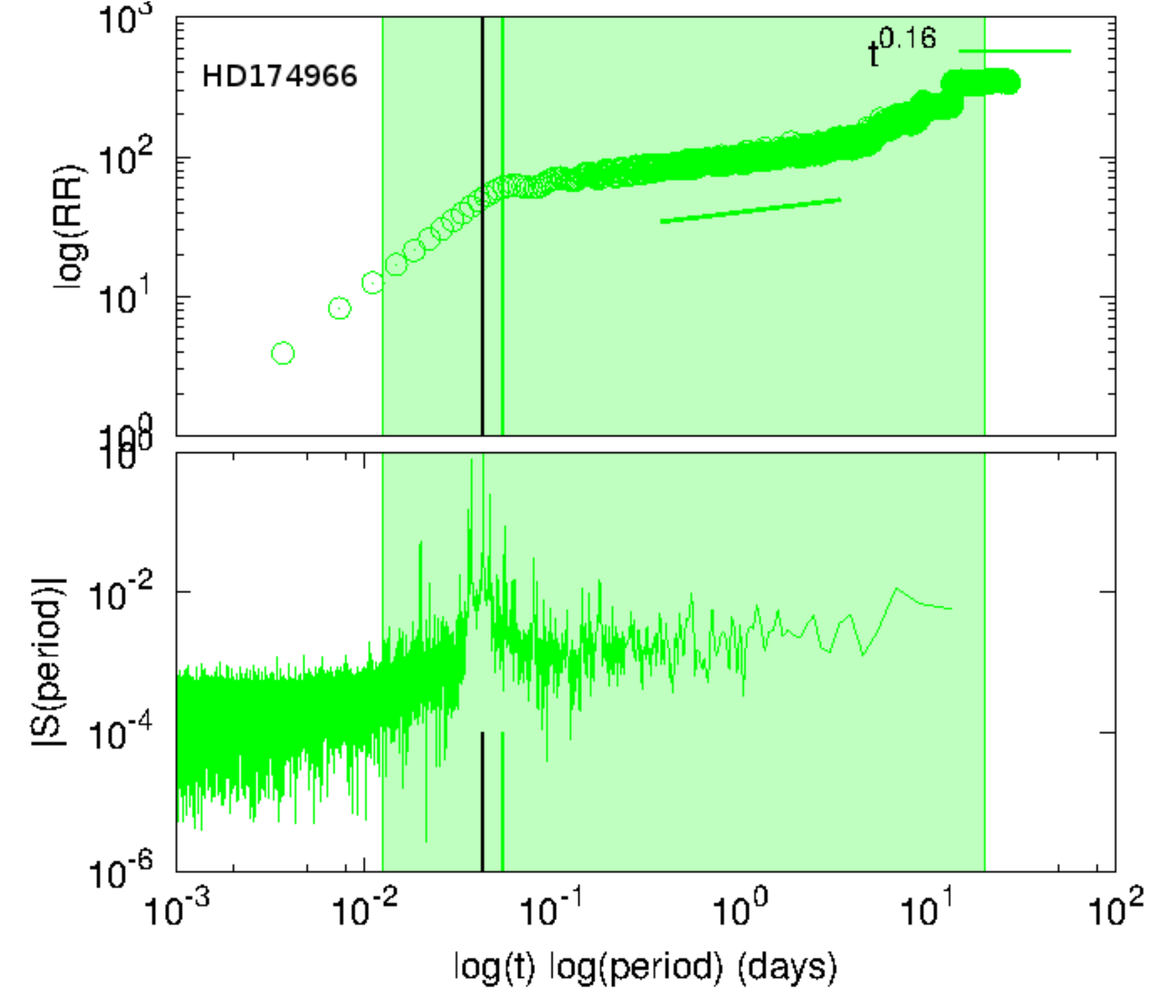}
    \includegraphics[width=8cm]{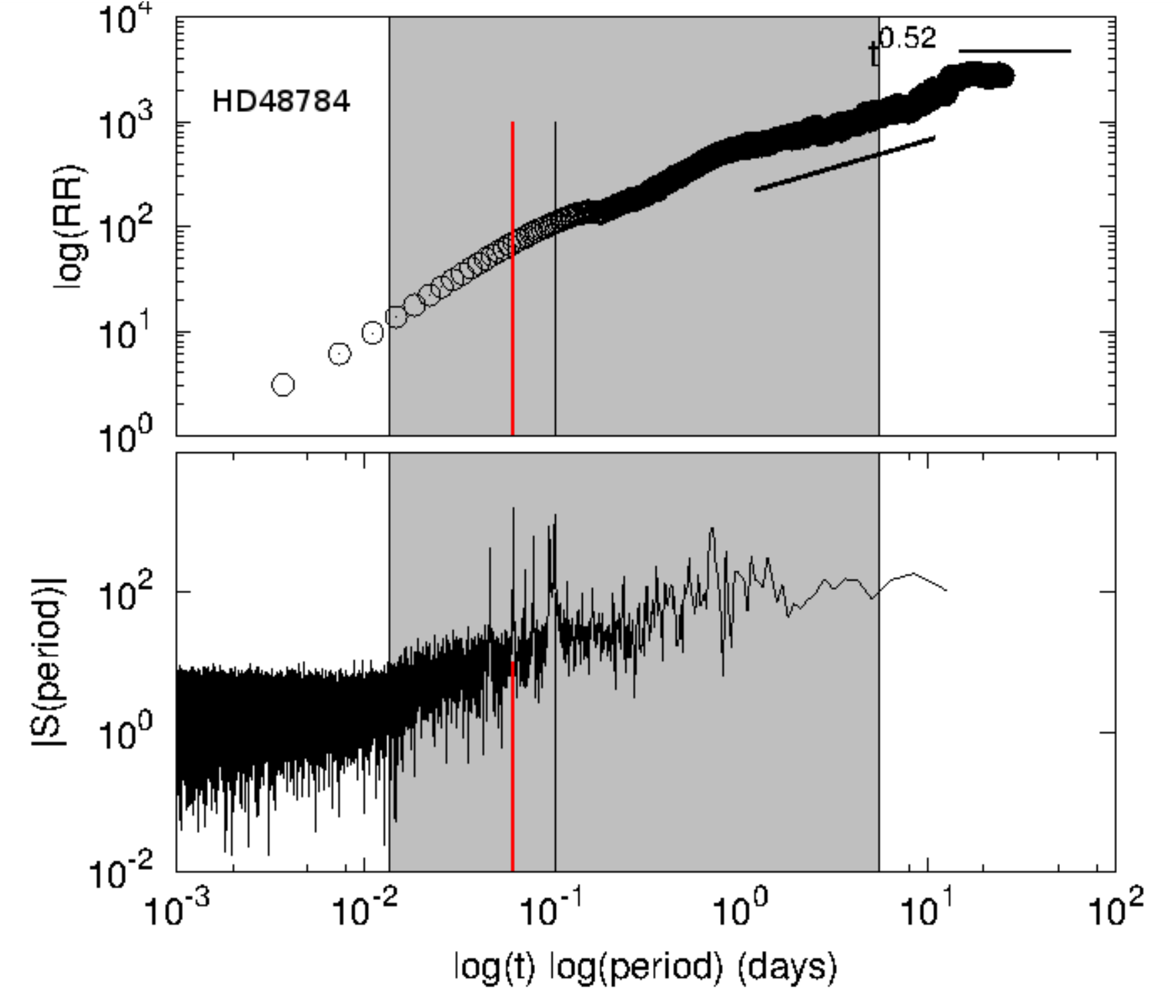}
    \includegraphics[width=8cm]{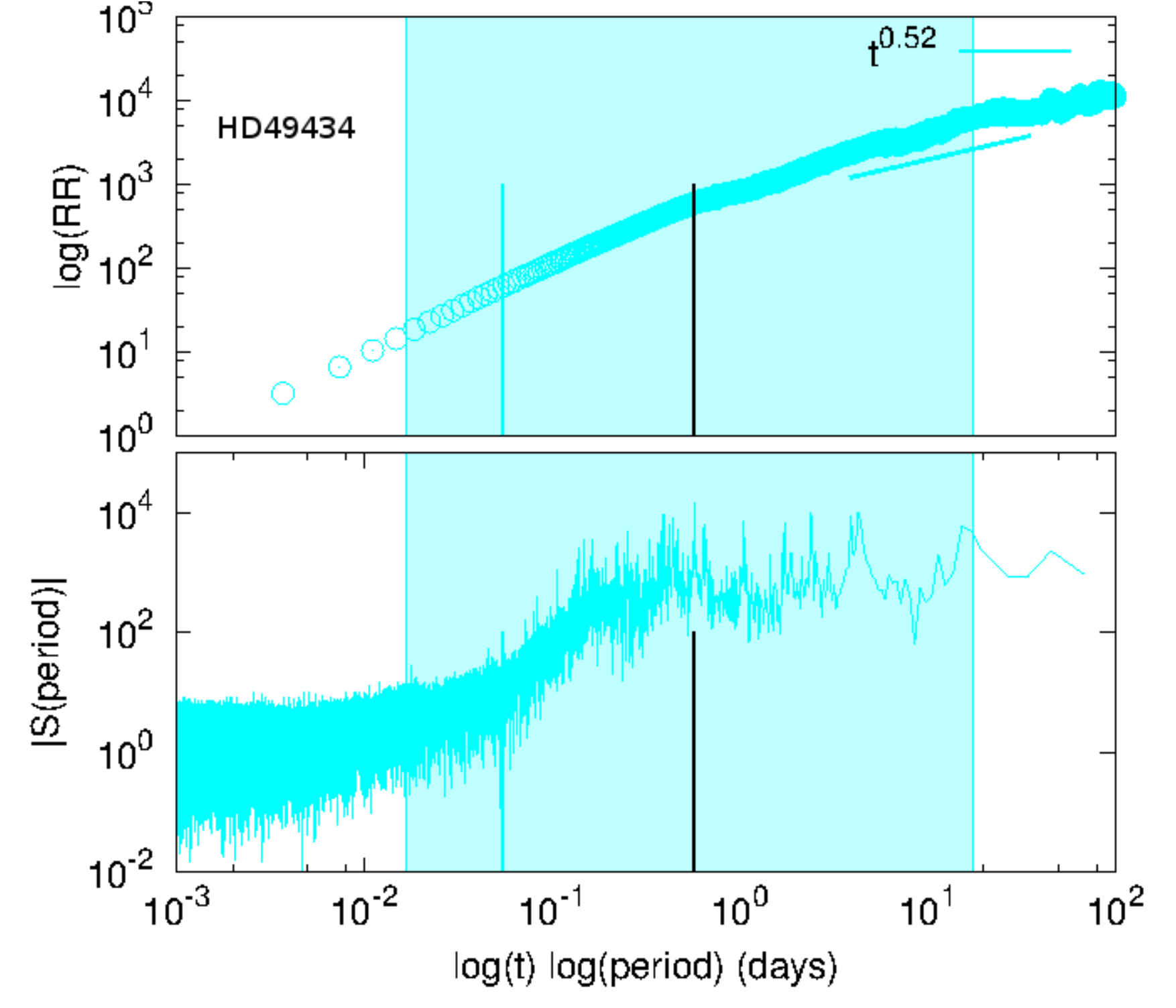}
    \includegraphics[width=8cm]{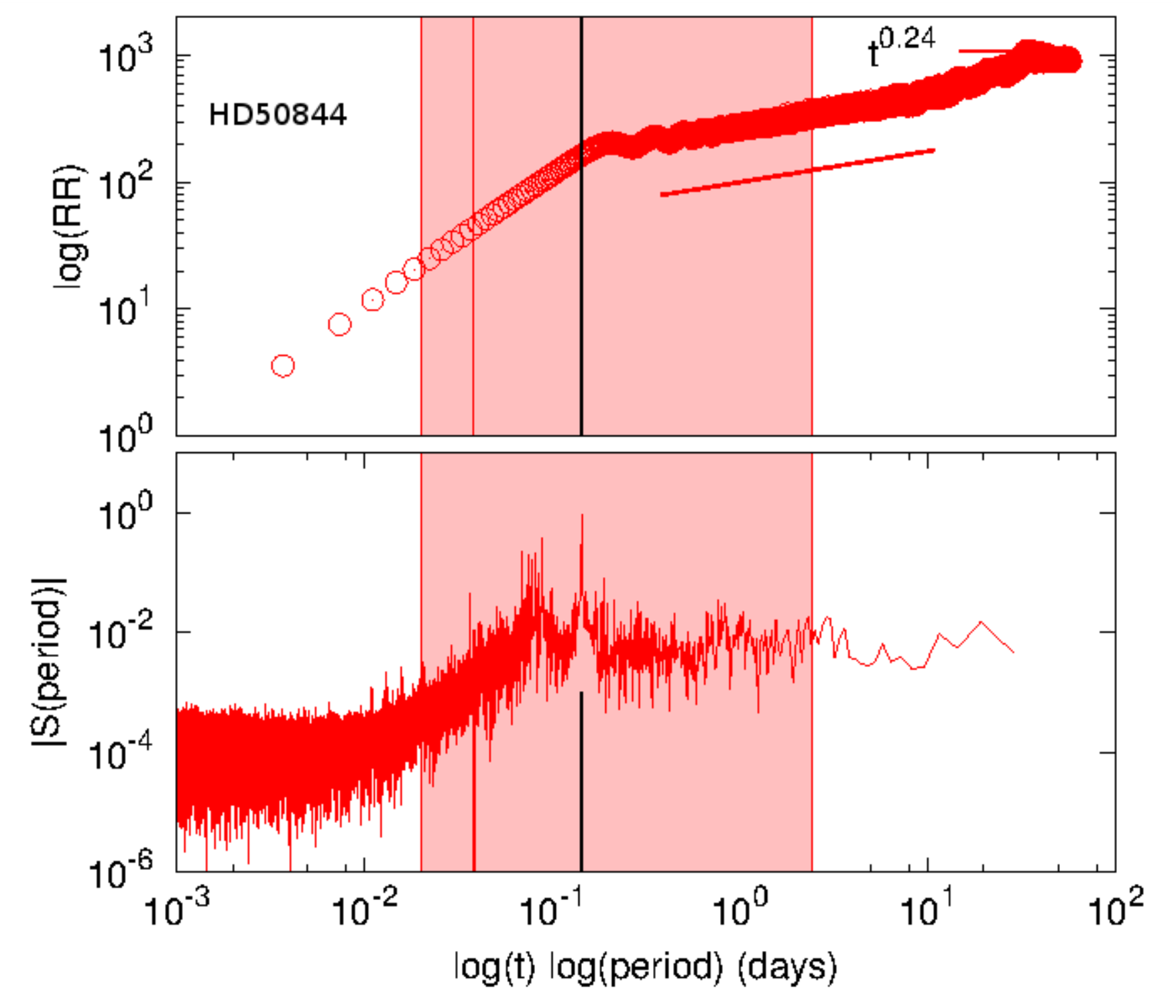}
	\includegraphics[width=8cm]{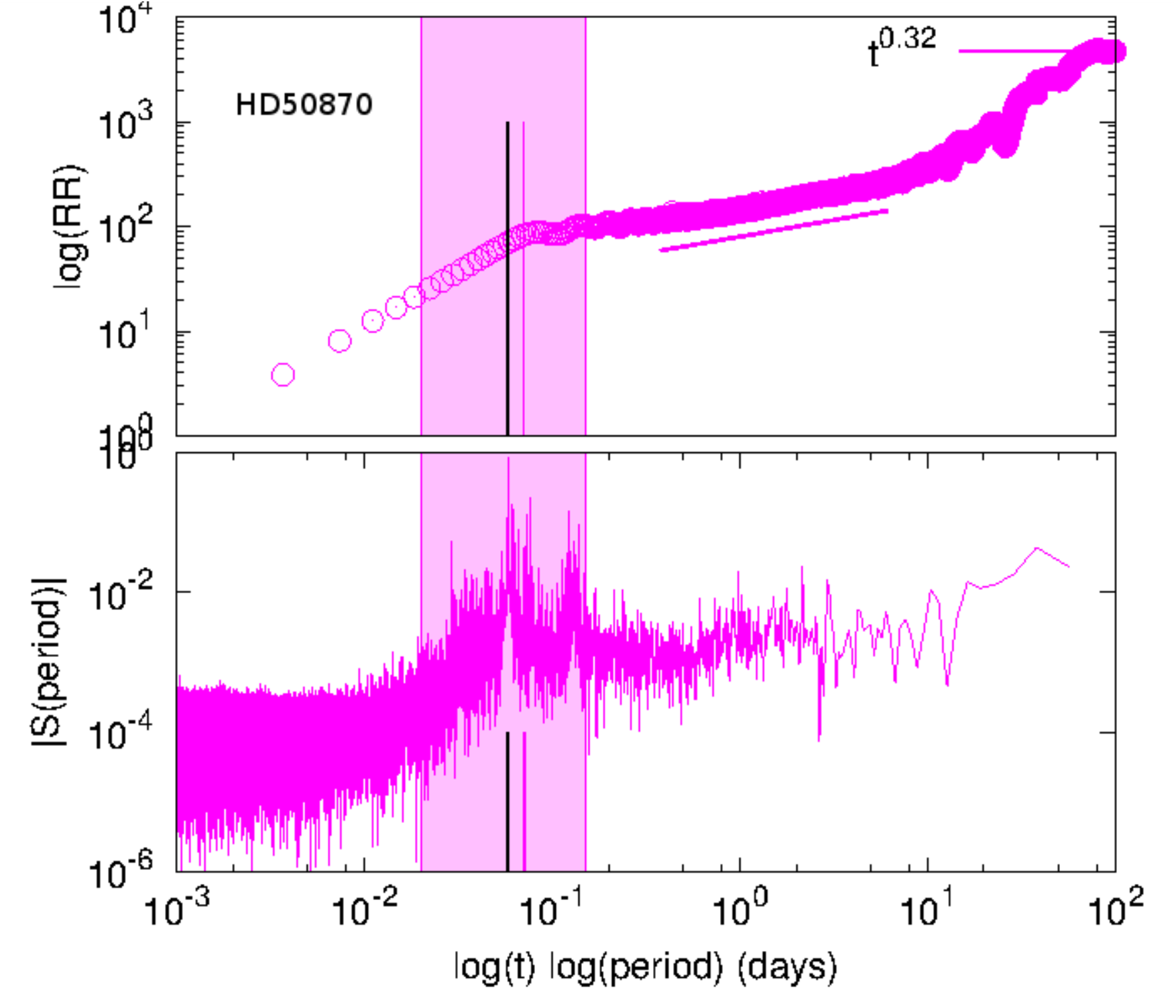}
	\caption{Rescaled Range and Power Spectra $|S|$ vs period for the 6 \dss\ analyzed in this study. Black and colored vertical lines corresponds to the period of $\nu_{max}$ and to the fundamental radial mode respectively. Shadowed band to the min-max period range of the modes detected in \protect\citet{Hernandez09HD174936, Garcia13HD174966,Javi18Prew,ChapellierHD49434,Poretti09HD5084, Mantegazza12HD50870} for HD\,174936, HD,174966, HD\,48784, HD\,49434, HD\,50844, and HD\,50870, respectively. Note that in log-log representations abscissa are inverted, i.e. $\log(\mathrm{frequency})=\log(\mathrm{period})^{-1}) = -\log(\mathrm{period})$.}
	\label{RR_all}
\end{figure*}
%
%
\section{Results and discussion}\label{sec:results}

%
%
%
%
%
%
%
%
The R/S analysis reveals a clear power law dependence with time for the 6 objects studied in this work (Figure~\ref{RR_dScuti}), thereby implying a strong evidence of the fractal nature of their light curves.  In addition, for each light curve we found at least two R/S log-log linear regimes, the first with slope $\alpha=1$ for $t< 10^{-1}$ days, and the second, with star characteristic dependent slope, emerges at $t\in(10^{-1},1)$ days and breaks down at $t\simeq 10$ days. In Figure~\ref{RR_all} (lower panels) we observe three different regimes: 
\begin{itemize}
	\item[$\bullet$] Short periods ($10^{-2}$ days). In the range of  minutes something similar to white noise is found.
    \item [$\bullet$] Medium periods ($10^{-2}:10^1$ days). From minutes and weeks, emerge the peaks that characterize the oscillation modes of \dss; these peaks are intertwined with one or more regions with power law behavior $\beta\sim 0$. Only in some cases such a power law have a clear positive slope, e.g. HD\,48784, which endures up to the final observation time.  Comparing upper and lower panels one can appreciate that the change in R/S slope corresponds to the period of fundamental radial mode, and in the second R/S linear regime we found both oscillating modes (included in the band of both panels) and a typical fractal regime, characterized by a noise similar to fractional Gaussian noise ($\alpha$ around $0.5$ and $\beta\sim 0$), with some indication of multifractality, i.e. $\beta\neq 2\alpha-1$ (see section \ref{sec:Spectra}). Some additional clues of the presence of multifractality are given by wavelet analysis, employed to study the Multifractal Singularity Spectra (see Figure~\ref{MF_dScuti} in Appendix~\ref{sec:MSS}).  We include the values of $\alpha$ and MSS width $\Delta\tilde{\alpha}$ in Table 2 for completeness. 
    \item[$\bullet$] High periods ($>10^1$ days). The power law behavior in this case could be the actual fingerprint of rotation and/or turbulence phenomena typical of the convective envelope.
\end{itemize}
%
%
The physical origin of the low periods fractal regime with R/S $\alpha=1$ and power spectra exponent $\beta=0$ is still unclear. Due to the short time range, data series are at most $270$ points long ($10^{-1}$ days with $32s$ of sampling), so R/S may be biased, and an extended study might be necessary, eventually with other statistical observables.
%
%
%
%

Our assumption is that CGSA is able to detect in the light curve only the contribution from stochastic, power-law-distributed, and linearly-generated noise, filtering out the contribution from oscillation modes, non-linear noise and chaotic dynamic. These results are in line with those of \citet{Yama}.\\
Exponents $\alpha$ seem to be in proportionality with $\fractal$ value, see Table~\ref{tab:CGSA_6Stars}, although further investigation with a larger dataset is needed.\\
%
%
In order to distinguish between the contributions to CGSA due to linearly generated noise and the ones coming from deterministic/stochastic nonlinear dynamic and low dimensional chaos, we analyzed the surrogated signals obtained by Fourier phase randomizing. This analysis has been previously developed and exploited for multifractal partition function studies \citep{Provenzale93} and correlation exponents \citep{Provenzale92},  so that if the correlation exponent converges also for the surrogate signal, then the origin of the convergence in the original time series cannot be related to its phase-space structure, and the hypothesis of nonlinear dynamic or low-dimensional chaos has to be rejected.  The same authors concluded that $\fractal$ of theoretical low-dimensional, non-linear, chaotic systems were not affected by Fourier phase randomizing.
\begin{figure}
	\centering
		\includegraphics[width=9cm]{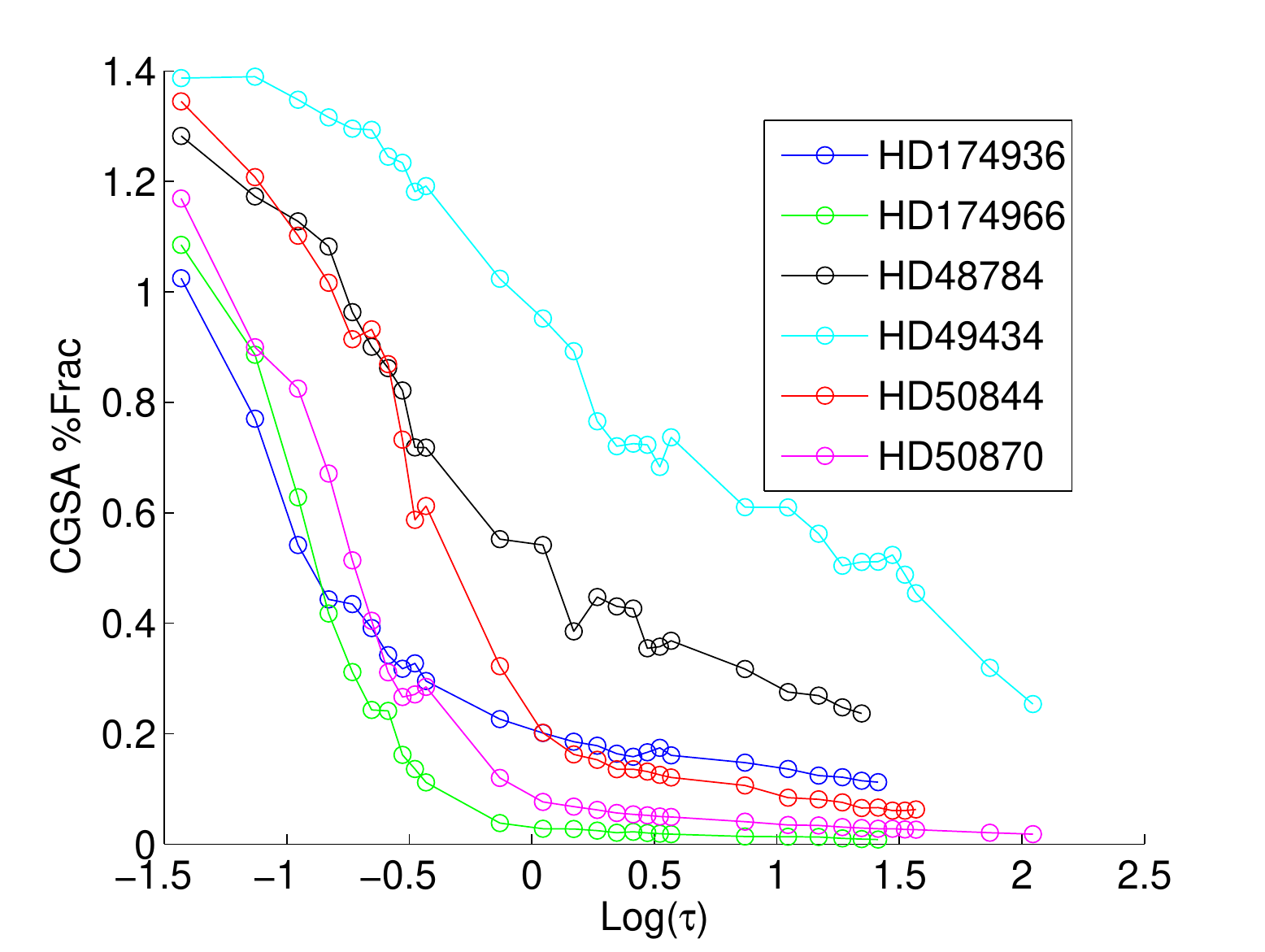}
	\caption{CGSA $\%Frac$ vs window size $\tau$ for the 6 Corot $\delta$-Scuti stars analyzed in this study. For each $\tau$ CGSA algorithm analyzed the $90\%$ of the total length, performing the cross-spectra orthogonalization among $N_{s}=50$ partially overlapping subsets.}
	\label{CGSA_vs_tau}
\end{figure}
%
%
%
%
%
Since light curves are composed by a superposition of deterministic phase-locked oscillators, regular and/or chaotic, and stochastic background, linear and/or nonlinear,  we used Fourier phase randomizing to separate linear stochastic component from the rest.  In most of the curves 	analyzed $\fractal \approx \fractalpr$, which could be a fingerprint of 	chaos and/or nonlinearity in the emergence of the signal\citep{Yama}.  On the contrary, the difference between $\fractal$ and the phase randomized $\fractalpr$ gives us an indication on the significance of power law noise in the whole signal.
%
%

%
\begin{figure*}
	\centering
	\includegraphics[width=7cm]{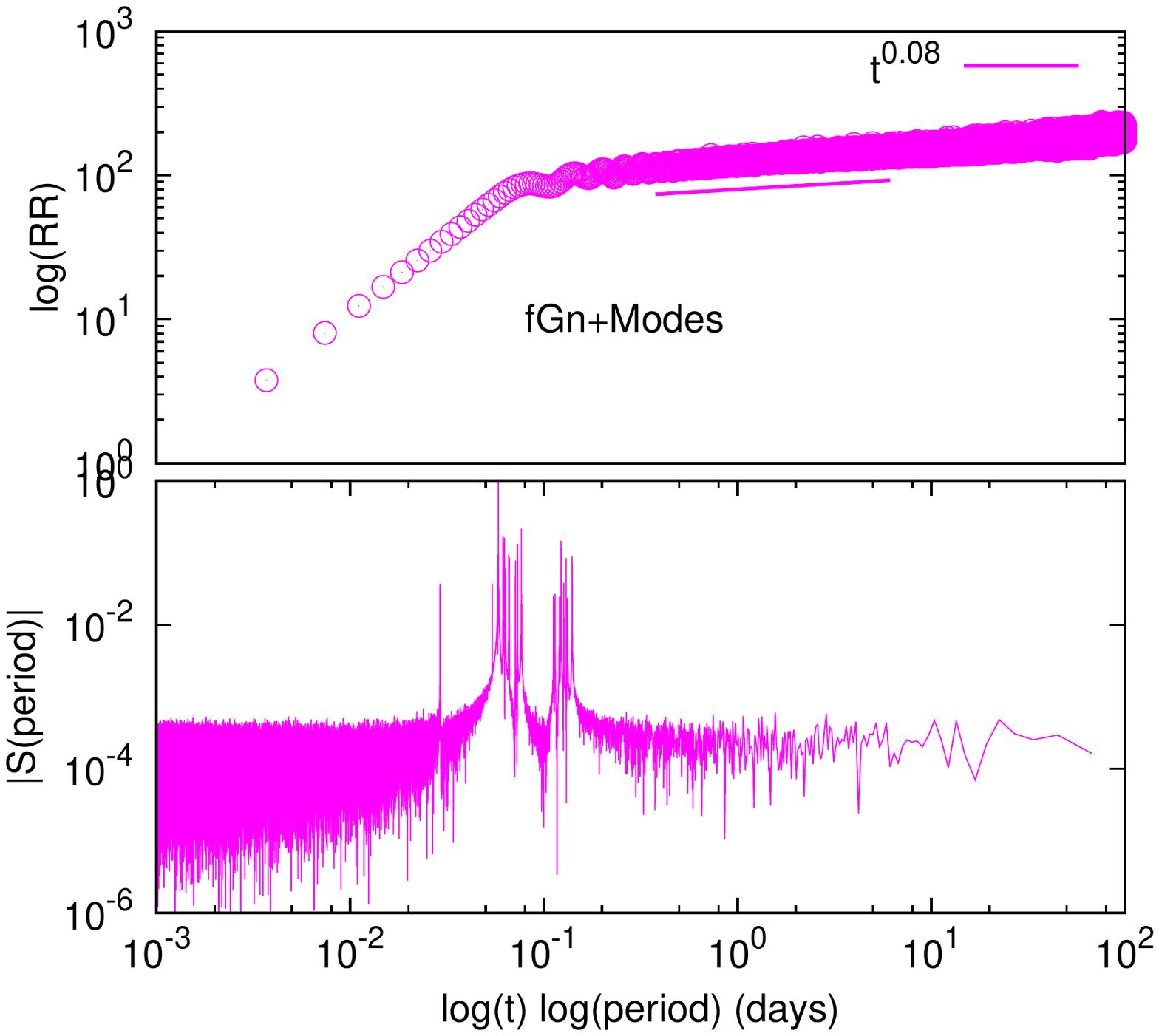}
    \includegraphics[width=7cm]{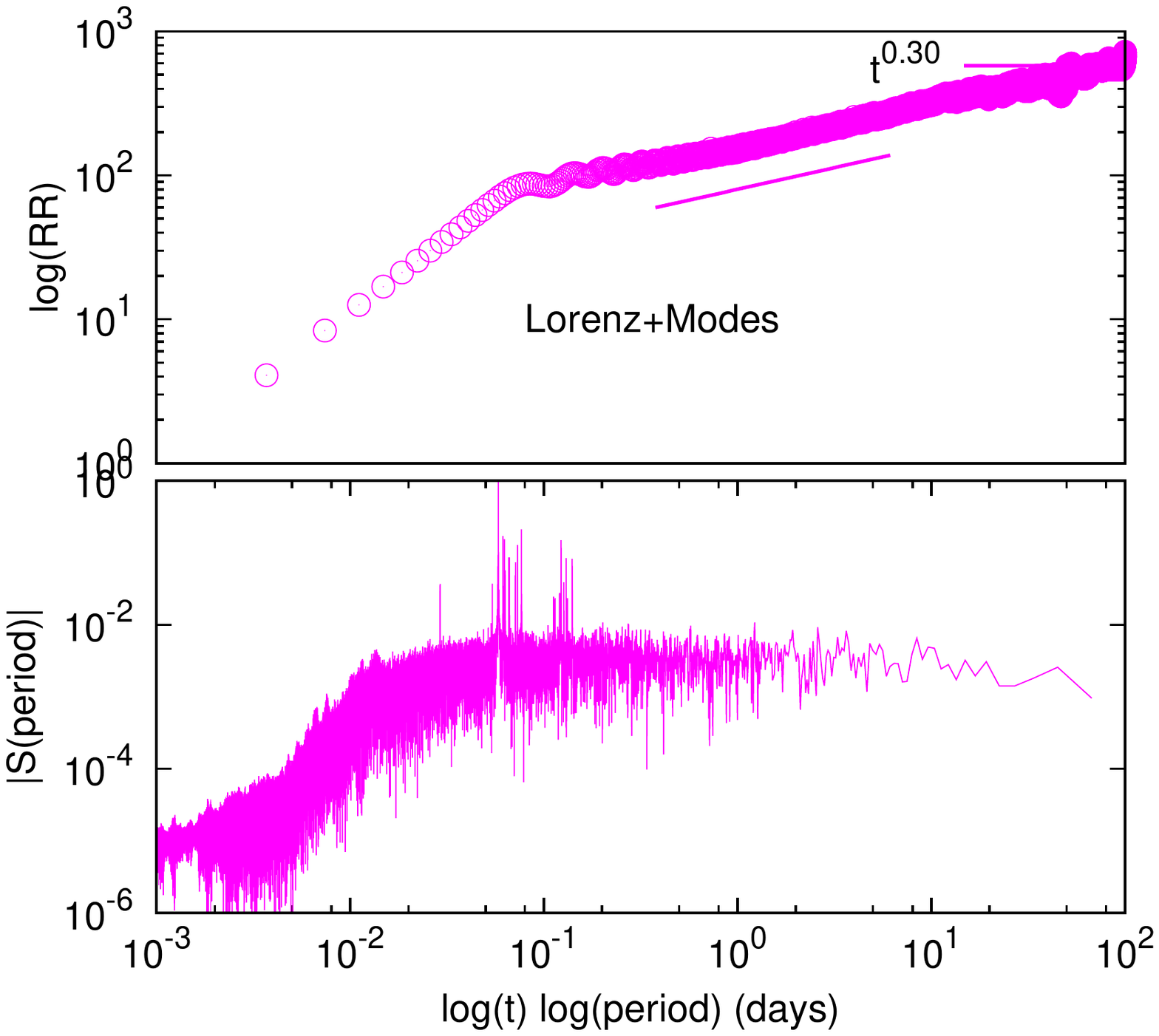}
    \includegraphics[width=7cm]{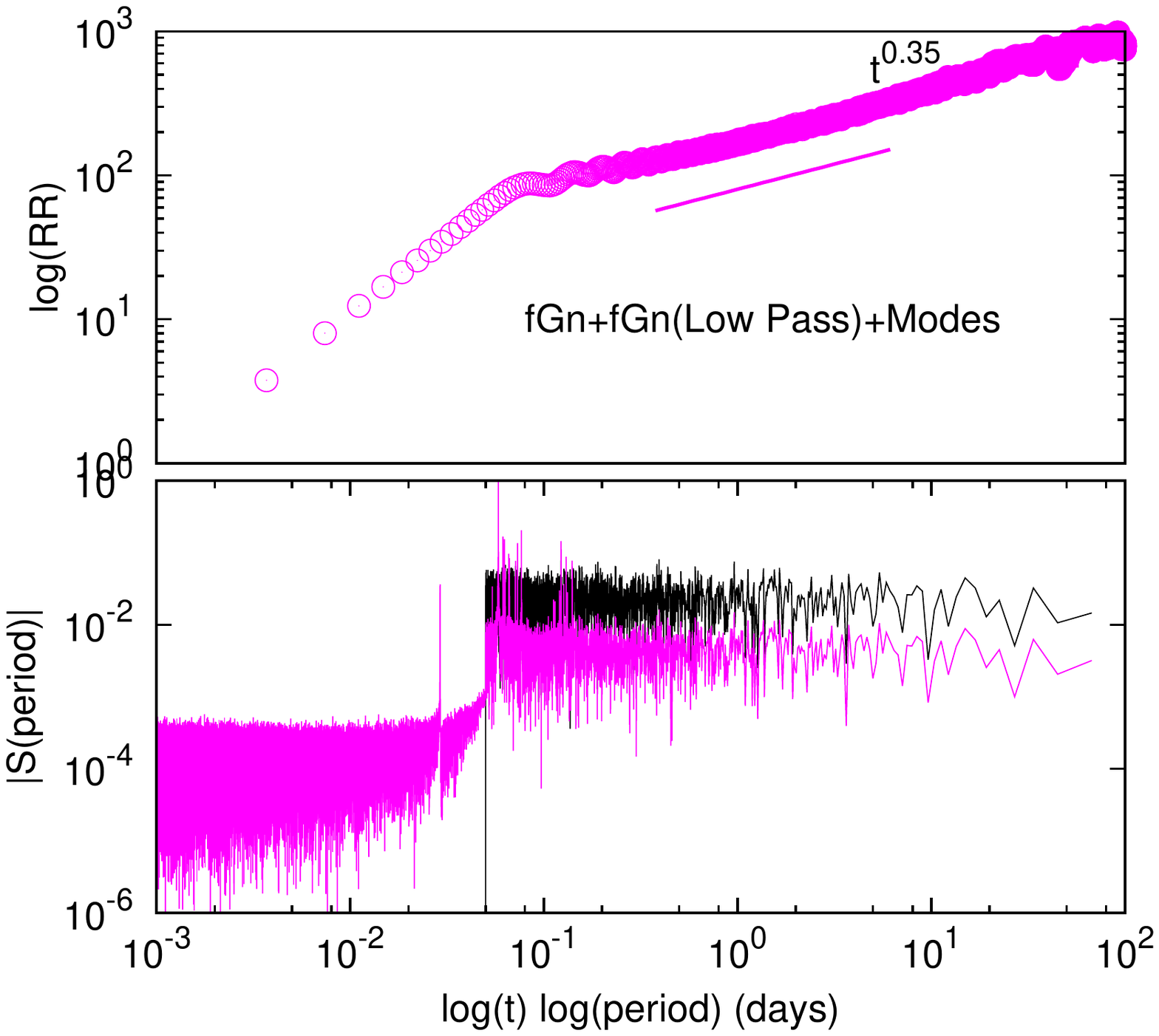}
     \includegraphics[width=7cm]{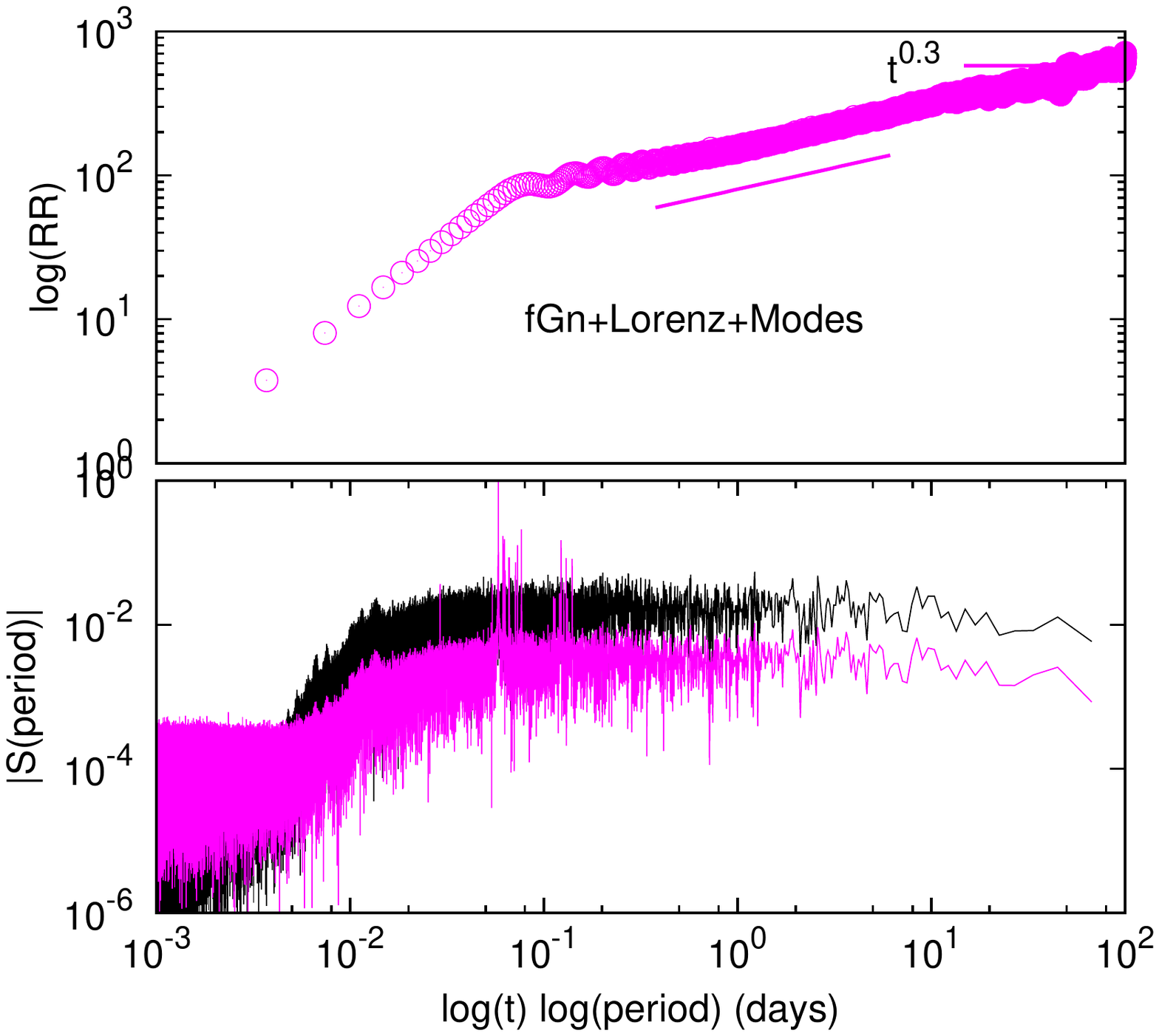}
	\caption{Rescaled Range and Power Spectra $|S|$ vs period for some toy models of HD50870. The 24 main oscillation modes have been taken from \protect\citep{Mantegazza12HD50870}. Upper left panel: background fGn with oscillation modes. Upper right panel: Lorenz model with oscillation modes. Lower left panel: sum of background fGn, Low Pass filtered fGn with higher amplitude, and oscillation modes. Lower right panel: sum of background fGn, Lorenz model with higher amplitude, and oscillation modes. Black curves in lower panels corresponds to fGn Low Pass (left) and Lorenz model power spectra (right).}\label{RR_Toy_HD50870}
\end{figure*}
In order to better understand how $\fractal$ can help in the analysis of modes and background noise, we study different temporal window sizes, i.e. we measure the mean value $<\fractal>_\tau$ over non-overlapping temporal windows with variable size $\tau$ (Figure~\ref{CGSA_vs_tau}). As expected, in all the curves $<\fractal>_\tau$ is decreasing with increasing window size $s$. This is because in larger windows are included more modes to the background, so that the total amount of fractality decreases, resulting eventually in a sort of sigmoidal curve. By exploiting this behavior one could develop a new criterion for selecting the temporal range where the modes can be computed \citep{Javi18Prew}, by finding lower and upper plateau in the curve. 
Interestingly, the two stars showing $\gamma$ Doradus pulsations (HD\,49434 and HD\,48784) show a quasi logarithmic decay (linear in log scale) whilst the remaining pure $\delta$ Scuti stars show a kind of linear decay. 
The $\delta$ Scuti stars here studied exhibit the highest amplitudes in a thin frequency region, which can be evaluated by mean of non-adiabatic contributions in the  $\kappa$ mechanism model \citep{Moya04}. On the other hand, the slower decaying of $\gamma$ Doradus may be due to a more homogeneous distribution of modes with quasi equal amplitude, similar to the background noise, see figure \ref{RR_all}.  Such modes are known to be excited by the blocking of convective flux \citep{Guzik2000}. 
This behavior, if confirmed with further studies, would provide a direct evidence for the presence of an external convective zone, since the convective blocking mechanism, responsible for maintaining  the $g$ modes, can only operate if the outer convective layer has a depth between 3 and 9 per cent of the stellar radius. It is worth assessing in future works if fractal analysis can constrain the size of such external convective envelope. This definitely would help to solve the mystery of the co-existence of pure $\delta$ Sct / $\gamma$ Dor and hybrid pulsators.

%
%
%
%
%
%

\section{Toy model for HD50870 fractal fingerprint}

We developed a toy model with the purpose to reproduce, at least qualitatively, the fractal fingerprint  of HD50870.  Let us focus on the power spectra of figure \ref{RR_Toy_HD50870}.  We modeled the flat spectra in low period (high frequency) domain with a Gaussian white noise $\eta(t)$ (see appendix \ref{sec:Frac_Models}). This the most simple signal producing a flat spectra. It is reasonable to choose it because $\fractal \sim 1$ in this regime. Then, we add the 24 oscillation modes studied in \citet{Mantegazza12HD50870}, i.e. $f(t)=\sum_{i=1}^{24}A_{i}\sin(\omega_it)$. This is shown in the upper left panel of Figure 4. There is a lack of any other information from R/S analysis and multifractal analysis; in fact the first $\alpha\sim 1$ R/S slope is due to the predominant amplitude of the main oscillation modes, and it is very hard to filter out the algorithm artifacts of multifractal analysis with temporal windows  of $10^{-2}$ days (27 points at most). This simple model produce the peak jumps in spectra around $period\approx 10^{-1}$. 
If we add an additional component which is the first variable $Lc(t)$ of Lorenz model \citep{Lorenz63}, the model reproduces the flat step in high period/low frequency spectra (see bottom right panel of Figure 4). With a proper choice of the relative amplitude for each signal the resulting synthetic signal is:
\begin{equation}
	S(t)=5\eta(t)+20Lc(t)+100f(t).
\end{equation}
Such signal resembles the fractal fingerprint of the data, as it is possible to appreciate in the good correspondence with R/S and power spectra analysis. Neglecting the term with $\eta(t)$ results in a more complex low period/high frequency spectra (see figure \ref{RR_Toy_HD50870} Up-Right panel), while without $Lc(t)$ in high period/low frequency regime the power spectra flat step disappears, and the slope of R/S  is almost flat (see figure \ref{RR_Toy_HD50870} Up-Left panel). It is possible to employ other kind of signals, deterministic or stochastic, to reproduce the high period/low frequency regime with a certain qualitative accuracy, for instance a low-pass filtered fGn, see figure \ref{RR_Toy_HD50870} Bottom-Left panel.  It is very important to go deep inside the essential properties and dynamics of high periods fluctuations, originated by granulation and/or rotational effects; in this context the definitive analysis could be the multifractal analysis in that regime.

The profile of $\fractal$ vs $t$ match very well the data for any synthetic model including  $f(t)$ (see Figure 5), suggesting that its decaying shape  is mainly dependent on the number, the amplitude and the frequency location of the pulsation modes.

However, the transition between low and high period regimes is smoother with a Lorenz model and resembles the results obtained with CoRoT data.

We have chosen HD50870 for the qualitative analysis through a toy model for simplicity but in our preliminary tests we obtained similar results for the rest of the stars of the set. Therefore, binarity or other specific characteristics of this object are not related to the results shown in this section.

\section{Conclusions and future prospects}\label{sec:conclusions}

This work aimed at detecting any fingerprint of fractal behavior in the light curves of \dss\ that may help to better understand their frequency content. 
To accomplish this objective we first undertook an investigation of the statistical techniques and algorithms better suited for this task, that is, those providing sufficient sensitivity to fractality at the different time scales present in the light curves of \dss. We came up with two: Rescaled Range Analysis and Coarse Graining Spectral Analysis, a Fourier Transform-based algorithm, which can discriminate stochastic fractal power spectra from the harmonic one. A third technique, Multifractal Singular Spectrum (MSS) was also applied but no conclusive results were found. Indeed, MSS has only tested so far in solar-like stars, which, in contrast with the \dss\, are mainly dominated by the stochastic phenomenon of convection (which also drives their oscillation modes). Even so, we found that MSS has a great potential for spotting  different fractal regimes, which will be exploited in future detailed works.

R/S analysis reveals a clear power law dependence with time for the 6 objects studied in this work, thereby implying an unambiguous detection of fractality. This analysis revealed three R/S regimes found systematically in all the objects. These regimes are: (1) short periods regime ($10^{-2}$ days), where fractal fingerprint of white noise was found; (2) medium periods regime ($10^{-2} - 10^1$ days), where evidences of multifractality were found in the region where the stellar modes pulsate, immediately after the main amplitude oscillating modes. Results are compatible with a combination of deterministic harmonics and a gaussian-like noise; (3) and high periods regime ($>10^1$ days), in which the origin is still unclear, although we speculate that this could be the actual fingerprint of rotation and/or turbulence phenomena typical of the convective envelope.

%
%
%
%
%

CGSA analysis was performed to filter out the contribution from the stellar oscillation modes, and in general of all chaotic-deterministic or stochastic nonlinear contributions, and thus isolate the fractal behavior of the stochastic, power-law-distributed, and linearly-generated noise. 
Using the Fourier phase randomizing technique to isolate the linear stochastic component of the signal we found that $\fractal$ measured on samples of phase randomized series is, in most of the cases, close to the original value, which might be a fingerprint of chaos and/or nonlinearity in the emergence of the signal. 

The analysis of CGSA variations with the temporal window yielded  a different behaviour for pure \dss\ and for those showing  $\gamma$ Doradus pulsations (HD\,49434 and HD\,48784). This surprising result may have a significant impact in the determination (and/or constrain) of the size of the external convective envelopes, which helps in solving the puzzling existence of pure $\delta$ Sct / $\gamma$ Dor and hybrid pulsators. 

The following steps are to apply this technique to a large sample of well-studied \dss, including hybrid pulsators (work in progress) with the objective of confirming the present results and finding relations between physical (and asteroseismic) observables with the different fractal parameters studied. Likewise, since we are able to spot the fractal behavior of the different components contributing to the light curves, we plan to apply fractal analysis as an unbiased criterion for detecting oscillation frequencies in the prewhitening process.  \\

%
\begin{figure}
	\centering
		\includegraphics[width=9cm]{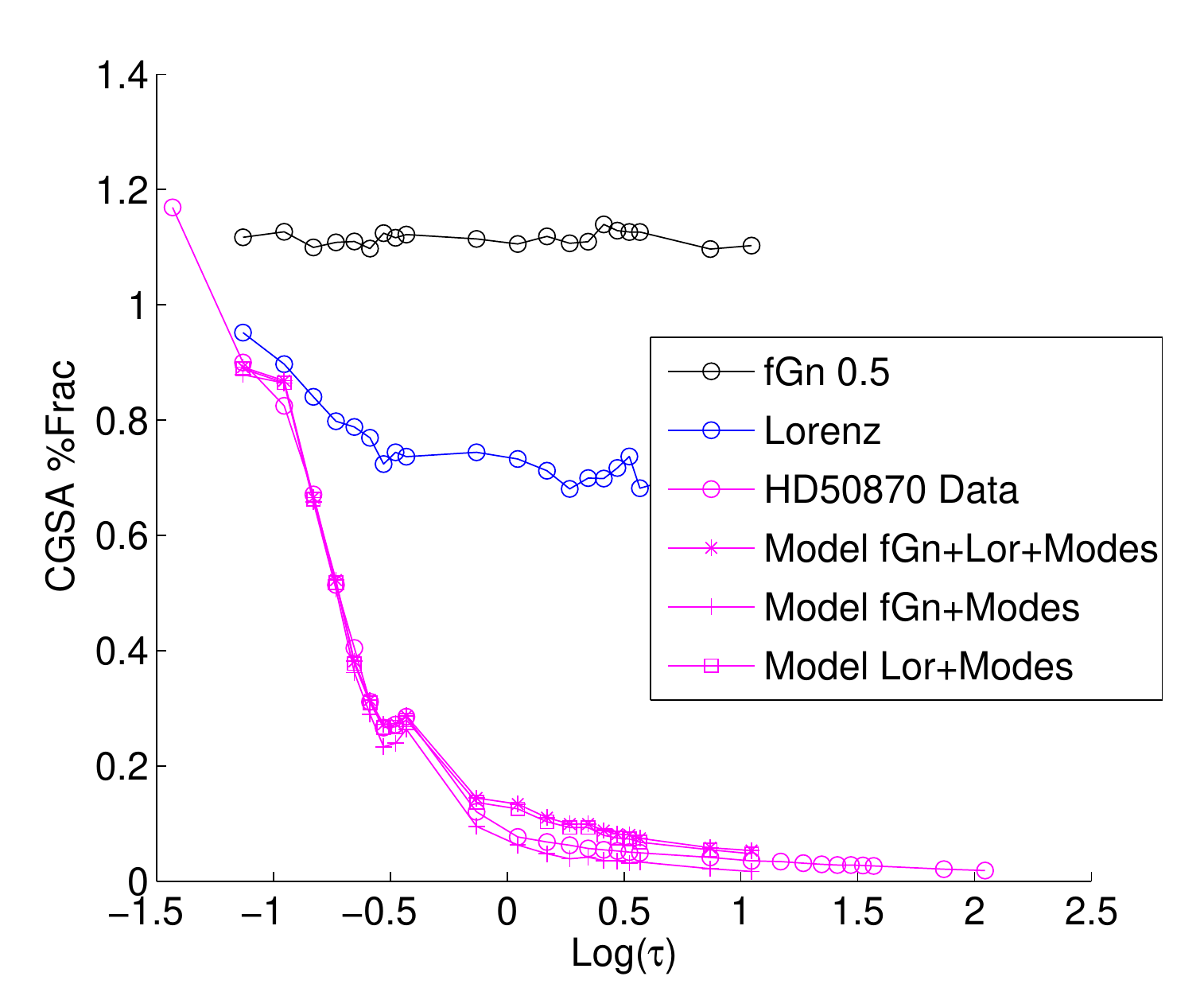}
	\caption{CGSA $\%Frac$ vs window size $\tau$ for fGn, Lorenz model, and data/models of HD50870.}
	\label{CGSA_vs_tau_Toy_HD50870}
\end{figure}
%

%

\section*{Acknowledgements}

SdF, JPG and RG acknowledge funding support from Spanish public funds for research under projects ESP2015-65712-C5-3-R. JCS acknowledges funding support from Spanish public funds for research under projects ESP2017-87676-2-2 and ESP2015-65712-C5-5-R, and from project RYC-2012-09913 under the 'Ram\'on y Cajal' program of the Spanish MINECO. SdF, JPG and RG acknowledge support from the 'Junta de Andaluc\'ia' local government under project 2012-P12-TIC-2469. Based on data from the COROT Archive at CAB.


\bibliographystyle{mnras}
\bibliography{lamiabib}

\begin{thebibliography}{}
\makeatletter
\relax
\def\mn@urlcharsother{\let\do\@makeother \do\$\do\&\do\#\do\^\do\_\do\%\do\~}
\def\mn@doi{\begingroup\mn@urlcharsother \@ifnextchar [ {\mn@doi@}
  {\mn@doi@[]}}
\def\mn@doi@[#1]#2{\def\@tempa{#1}\ifx\@tempa\@empty \href
  {http://dx.doi.org/#2} {doi:#2}\else \href {http://dx.doi.org/#2} {#1}\fi
  \endgroup}
\def\mn@eprint#1#2{\mn@eprint@#1:#2::\@nil}
\def\mn@eprint@arXiv#1{\href {http://arxiv.org/abs/#1} {{\tt arXiv:#1}}}
\def\mn@eprint@dblp#1{\href {http://dblp.uni-trier.de/rec/bibtex/#1.xml}
  {dblp:#1}}
\def\mn@eprint@#1:#2:#3:#4\@nil{\def\@tempa {#1}\def\@tempb {#2}\def\@tempc
  {#3}\ifx \@tempc \@empty \let \@tempc \@tempb \let \@tempb \@tempa \fi \ifx
  \@tempb \@empty \def\@tempb {arXiv}\fi \@ifundefined
  {mn@eprint@\@tempb}{\@tempb:\@tempc}{\expandafter \expandafter \csname
  mn@eprint@\@tempb\endcsname \expandafter{\@tempc}}}

\bibitem[\protect\citeauthoryear{Antoci et~al.,}{Antoci
  et~al.}{2011}]{Antoci11}
Antoci V.,  et~al., 2011, Nature, 477, 570

\bibitem[\protect\citeauthoryear{{Aschwanden}}{{Aschwanden}}{2012}]{SOCAstro12}
{Aschwanden} M.~J.,  2012, preprint (\mn@eprint {arXiv} {1207.4413})

\bibitem[\protect\citeauthoryear{Baglin et~al.,}{Baglin
  et~al.}{2006}]{Baglin06}
Baglin A.,  et~al., 2006, in 36th COSPAR Scientific assembly.

\bibitem[\protect\citeauthoryear{{Balona}}{{Balona}}{2011}]{Balona11}
{Balona} L.~A.,  2011, \mn@doi [\mnras] {10.1111/j.1365-2966.2011.18813.x},
  \href {https://ui.adsabs.harvard.edu/#abs/2011MNRAS.415.1691B} {415, 1691}

\bibitem[\protect\citeauthoryear{Balona, Daszy\`nska-Daszkiewicz  \&
  Pamyatnykh}{Balona et~al.}{2015}]{Balona2015}
Balona L.~A.,  Daszy\`nska-Daszkiewicz J.,   Pamyatnykh A.~A.,  2015, \mn@doi
  [Monthly Notices of the Royal Astronomical Society] {10.1093/mnras/stv1513},
  452, 3073

\bibitem[\protect\citeauthoryear{Barab{\'a}si \& Stanley}{Barab{\'a}si \&
  Stanley}{1995}]{BarabasiBook95}
Barab{\'a}si A.-L.,  Stanley H.~E.,  1995, Fractal concepts in surface growth.
Cambridge university press

\bibitem[\protect\citeauthoryear{{Barcel{\'o} Forteza}, {Michel}, {Roca
  Cort{\'e}s}  \& {Garc{\'{\i}}a}}{{Barcel{\'o} Forteza}
  et~al.}{2015}]{BarceloForteza15}
{Barcel{\'o} Forteza} S.,  {Michel} E.,  {Roca Cort{\'e}s} T.,
  {Garc{\'{\i}}a} R.~A.,  2015, \mn@doi [\aap] {10.1051/0004-6361/201425507},
  \href {http://adsabs.harvard.edu/abs/2015A%26A...579A.133B} {579, A133}

\bibitem[\protect\citeauthoryear{{Barcel{\'o} Forteza}, {Roca Cort{\'e}s},
  {Garc{\'{\i}}a Hern{\'a}ndez}  \& {Garc{\'{\i}}a}}{{Barcel{\'o} Forteza}
  et~al.}{2017}]{BarceloForteza17}
{Barcel{\'o} Forteza} S.,  {Roca Cort{\'e}s} T.,  {Garc{\'{\i}}a Hern{\'a}ndez}
  A.,   {Garc{\'{\i}}a} R.~A.,  2017, \mn@doi [\aap]
  {10.1051/0004-6361/201628675}, \href
  {http://adsabs.harvard.edu/abs/2017A%26A...601A..57B} {601, A57}

\bibitem[\protect\citeauthoryear{Benzi, Paladin, Parisi  \& Vulpiani}{Benzi
  et~al.}{1984}]{Benzi84}
Benzi R.,  Paladin G.,  Parisi G.,   Vulpiani A.,  1984, Journal of Physics A:
  Mathematical and General, 17, 3521

\bibitem[\protect\citeauthoryear{Chandler}{Chandler}{1987}]{Chandler87}
Chandler D.,  1987, Introduction to Modern Statistical Mechanics, by David
  Chandler, pp. 288. Foreword by David Chandler. Oxford University Press, Sep
  1987. ISBN-10: 0195042778. ISBN-13: 9780195042771, p.~288

\bibitem[\protect\citeauthoryear{{Chapellier, E.} et~al.,}{{Chapellier, E.}
  et~al.}{2011}]{ChapellierHD49434}
{Chapellier, E.} et~al., 2011, Astronomy \& Astrophysics, 525, A23

\bibitem[\protect\citeauthoryear{Charbonneau, McIntosh, Liu  \&
  Bogdan}{Charbonneau et~al.}{2001}]{Charbo01}
Charbonneau P.,  McIntosh S.~W.,  Liu H.-L.,   Bogdan T.~J.,  2001, Solar
  Physics, 203, 321

\bibitem[\protect\citeauthoryear{Cheng}{Cheng}{2014}]{Cheng14}
Cheng Q.,  2014, \mn@doi [Nonlinear Processes in Geophysics]
  {10.5194/npg-21-477-2014}, 21, 477

\bibitem[\protect\citeauthoryear{De~Bartolo, Gabriele  \& Gaudio}{De~Bartolo
  et~al.}{2000}]{DeBartolo00}
De~Bartolo S.~G.,  Gabriele S.,   Gaudio R.,  2000, {Hydrology and Earth System
  Sciences Discussions}, 4, 105

\bibitem[\protect\citeauthoryear{{De Freitas}, {Le{\~a}o}, {Ferreira Lopes},
  {Paz-Chinchon}, {Canto Martins}, {Alves}, {De Medeiros}  \& {Catelan}}{{De
  Freitas} et~al.}{2013}]{NewSunsI}
{De Freitas} D.~B.,  {Le{\~a}o} I.~C.,  {Ferreira Lopes} C.~E.,  {Paz-Chinchon}
  F.,  {Canto Martins} B.~L.,  {Alves} S.,  {De Medeiros} J.~R.,   {Catelan}
  M.,  2013, \mn@doi [\apjl] {10.1088/2041-8205/773/2/L18}, \href
  {http://adsabs.harvard.edu/abs/2013ApJ...773L..18D} {773, L18}

\bibitem[\protect\citeauthoryear{{De Freitas} et~al.,}{{De Freitas}
  et~al.}{2016}]{NewSunsIII}
{De Freitas} D.~B.,  et~al., 2016, \mn@doi [\apj] {10.3847/0004-637X/831/1/87},
  \href {http://adsabs.harvard.edu/abs/2016ApJ...831...87D} {831, 87}

\bibitem[\protect\citeauthoryear{{De Freitas}, {Nepomuceno}, {Gomes de Souza},
  {Leão}, {Das Chagas}, {Costa}, {Canto Martins}  \& {De Medeiros}}{{De
  Freitas} et~al.}{2017}]{NewSunsIV}
{De Freitas} D.~B.,  {Nepomuceno} M.~M.~F.,  {Gomes de Souza} M.,  {Leão}
  I.~C.,  {Das Chagas} M.~L.,  {Costa} A.~D.,  {Canto Martins} B.,   {De
  Medeiros} J.~R.,  2017, The Astrophysical Journal, 843, 103

\bibitem[\protect\citeauthoryear{Dickman, Mu{\~n}oz, Vespignani  \&
  Zapperi}{Dickman et~al.}{2000}]{Dickman00}
Dickman R.,  Mu{\~n}oz M.~A.,  Vespignani A.,   Zapperi S.,  2000, Brazilian
  Journal of Physics, 30, 27

\bibitem[\protect\citeauthoryear{Dom{\'\i}nguez-Tenreiro, Roy  \&
  Martinez}{Dom{\'\i}nguez-Tenreiro et~al.}{1992}]{Dominguez92}
Dom{\'\i}nguez-Tenreiro R.,  Roy L.,   Martinez V.,  1992, Progress of
  theoretical physics, 87, 1107

\bibitem[\protect\citeauthoryear{Drozdz \& Oswiecimka}{Drozdz \&
  Oswiecimka}{2015}]{drozdz}
Drozdz S.,  Oswiecimka P.,  2015, Physical Review E, 91, 030902

\bibitem[\protect\citeauthoryear{Garc{\'\i}a~Hern{\'a}ndez
  et~al.,}{Garc{\'\i}a~Hern{\'a}ndez et~al.}{2009}]{Hernandez09HD174936}
Garc{\'\i}a~Hern{\'a}ndez A.,  et~al., 2009, Astronomy \& Astrophysics, 506, 79

\bibitem[\protect\citeauthoryear{Garc{\'\i}a~Hern{\'a}ndez
  et~al.,}{Garc{\'\i}a~Hern{\'a}ndez et~al.}{2013}]{Garcia13HD174966}
Garc{\'\i}a~Hern{\'a}ndez A.,  et~al., 2013, Astronomy \&
  Astrophysics/Astronomie et Astrophysique, 559

\bibitem[\protect\citeauthoryear{Gilliland et~al.,}{Gilliland
  et~al.}{2010}]{Gilli10}
Gilliland R.~L.,  et~al., 2010, The Astrophysical Journal Letters, 713, L160

\bibitem[\protect\citeauthoryear{Goldreich \& Keeley}{Goldreich \&
  Keeley}{1977}]{Gold77}
Goldreich P.,  Keeley D.,  1977, The Astrophysical Journal, 212, 243

\bibitem[\protect\citeauthoryear{{Goupil}, {Dupret}, {Samadi}, {Boehm},
  {Alecian}, {Suarez}, {Lebreton}  \& {Catala}}{{Goupil}
  et~al.}{2005}]{Goupil05}
{Goupil} M.-J.,  {Dupret} M.~A.,  {Samadi} R.,  {Boehm} T.,  {Alecian} E.,
  {Suarez} J.~C.,  {Lebreton} Y.,   {Catala} C.,  2005, \mn@doi [Journal of
  Astrophysics and Astronomy] {10.1007/BF02702333}, \href
  {http://adsabs.harvard.edu/abs/2005JApA...26..249G} {26, 249}

\bibitem[\protect\citeauthoryear{{Grigahc{\`e}ne} et~al.,}{{Grigahc{\`e}ne}
  et~al.}{2010}]{Ahmed2010}
{Grigahc{\`e}ne} A.,  et~al., 2010, \mn@doi [\apjl]
  {10.1088/2041-8205/713/2/L192}, \href
  {http://adsabs.harvard.edu/abs/2010ApJ...713L.192G} {713, L192}

\bibitem[\protect\citeauthoryear{Guzik, Kaye, Bradley, Cox  \& Neuforge}{Guzik
  et~al.}{2000}]{Guzik2000}
Guzik J.~A.,  Kaye A.~B.,  Bradley P.~A.,  Cox A.~N.,   Neuforge C.,  2000, The
  Astrophysical Journal Letters, 542, L57

\bibitem[\protect\citeauthoryear{Hnat, Chapman, Kiyani, Rowlands  \&
  Watkins}{Hnat et~al.}{2007}]{Chapman07}
Hnat B.,  Chapman S.~C.,  Kiyani K.,  Rowlands G.,   Watkins N.~W.,  2007,
  Geophysical Research Letters, 34

\bibitem[\protect\citeauthoryear{{Kallinger} \& {Matthews}}{{Kallinger} \&
  {Matthews}}{2010}]{KallingerMat10}
{Kallinger} T.,  {Matthews} J.~M.,  2010, \mn@doi [\apj]
  {10.1088/2041-8205/711/1/L35}, \href
  {https://ui.adsabs.harvard.edu/#abs/2010ApJ...711L..35K} {711, L35}

\bibitem[\protect\citeauthoryear{Kantelhardt}{Kantelhardt}{2008}]{Kantel08}
Kantelhardt J.~W.,  2008, {Fractal and Multifractal Time Series} (\mn@eprint {}
  {0804.0747}), \url {http://arxiv.org/abs/0804.0747}

\bibitem[\protect\citeauthoryear{{Lemmerer}, {Hanslmeier}, {Muthsam}  \&
  {Piantschitsch}}{{Lemmerer} et~al.}{2017}]{Lemme17}
{Lemmerer} B.,  {Hanslmeier} A.,  {Muthsam} H.,   {Piantschitsch} I.,  2017,
  \mn@doi [\aap] {10.1051/0004-6361/201528011}, \href
  {http://adsabs.harvard.edu/abs/2017A%26A...598A.126L} {598, A126}

\bibitem[\protect\citeauthoryear{{Ligni{\`e}res} \& {Georgeot}}{{Ligni{\`e}res}
  \& {Georgeot}}{2009}]{LignieresGeor09}
{Ligni{\`e}res} F.,  {Georgeot} B.,  2009, \mn@doi [\aap]
  {10.1051/0004-6361/200811165}, \href
  {https://ui.adsabs.harvard.edu/#abs/2009A&A...500.1173L} {500, 1173}

\bibitem[\protect\citeauthoryear{Lindner, Kohar, Kia, Hippke, Learned  \&
  Ditto}{Lindner et~al.}{2015}]{lindner15}
Lindner J.~F.,  Kohar V.,  Kia B.,  Hippke M.,  Learned J.~G.,   Ditto W.~L.,
  2015, Physical review letters, 114, 054101

\bibitem[\protect\citeauthoryear{Lorenz}{Lorenz}{1963}]{Lorenz63}
Lorenz E.~N.,  1963, Journal of the atmospheric sciences, 20, 130

\bibitem[\protect\citeauthoryear{Lyra \& Tsallis}{Lyra \&
  Tsallis}{1998}]{Lyra98}
Lyra M.,  Tsallis C.,  1998, Physical review letters, 80, 53

\bibitem[\protect\citeauthoryear{Malamud \& Turcotte}{Malamud \&
  Turcotte}{1999}]{Turco}
Malamud B.~D.,  Turcotte D.~L.,  1999, \mn@doi [Journal of Statistical Planning
  and Inference] {https://doi.org/10.1016/S0378-3758(98)00249-3}, 80, 173

\bibitem[\protect\citeauthoryear{Mandelbrot}{Mandelbrot}{1977}]{manda77}
Mandelbrot B.,  1977, The fractal geometry of nature.
Freeman, \url {https://books.google.es/books?id=JFX9mQEACAAJ}

\bibitem[\protect\citeauthoryear{Mandelbrot \& Scholz}{Mandelbrot \&
  Scholz}{1989}]{Scholz89}
Mandelbrot B.~B.,  Scholz C.~H.,  1989, Fractals in geophysics.
Birkhauser Verlag

\bibitem[\protect\citeauthoryear{Mantegazza et~al.,}{Mantegazza
  et~al.}{2012}]{Mantegazza12HD50870}
Mantegazza L.,  et~al., 2012, Astronomy \& Astrophysics, 542, A24

\bibitem[\protect\citeauthoryear{Mathur, Salabert, Garc{\'\i}a  \&
  Ceillier}{Mathur et~al.}{2014}]{mathur2014photometric}
Mathur S.,  Salabert D.,  Garc{\'\i}a R.~A.,   Ceillier T.,  2014, Journal of
  Space Weather and Space Climate, 4, A15

\bibitem[\protect\citeauthoryear{Meakin \& Jamtveit}{Meakin \&
  Jamtveit}{2010}]{Meakin09}
Meakin P.,  Jamtveit B.,  2010, \mn@doi [Proceedings of the Royal Society of
  London A] {10.1098/rspa.2009.0189}, 466, 659

\bibitem[\protect\citeauthoryear{Meneveau \& Sreenivasan}{Meneveau \&
  Sreenivasan}{1987}]{Meneveau87}
Meneveau C.,  Sreenivasan K.,  1987, Physical review letters, 59, 1424

\bibitem[\protect\citeauthoryear{{Moya}, {Garrido}  \& {Dupret}}{{Moya}
  et~al.}{2004}]{Moya04}
{Moya} A.,  {Garrido} R.,   {Dupret} M.~A.,  2004, \aap, 414, 1081

\bibitem[\protect\citeauthoryear{{Neiner} \& {Lampens}}{{Neiner} \&
  {Lampens}}{2015}]{NeinerLampens15}
{Neiner} C.,  {Lampens} P.,  2015, \mn@doi [\mnras] {10.1093/mnrasl/slv130},
  \href {https://ui.adsabs.harvard.edu/#abs/2015MNRAS.454L..86N} {454, L86}

\bibitem[\protect\citeauthoryear{{Pascual-Granado}}{{Pascual-Granado}}{2011}]{JaviFrac11}
{Pascual-Granado} J.,  2011, in {Zapatero Osorio} M.~R.,  {Gorgas} J.,
  {Ma{\'{\i}}z Apell{\'a}niz} J.,  {Pardo} J.~R.,   {Gil de Paz} A.,  eds,
  Highlights of Spanish Astrophysics VI. pp 744--748

\bibitem[\protect\citeauthoryear{Pascual-Granado, Garrido  \&
  Su{\'a}rez}{Pascual-Granado et~al.}{2015}]{Javi15miarma}
Pascual-Granado J.,  Garrido R.,   Su{\'a}rez J.,  2015, Astronomy \&
  Astrophysics, 575, A78

\bibitem[\protect\citeauthoryear{Pascual-Granado, Su\'arez, Garrido, Moya,
  Garc\'ia~Hern\'andez, Rod\'on  \& Lares-Martiz}{Pascual-Granado
  et~al.}{2018}]{Javi18Prew}
Pascual-Granado J.,  Su\'arez J.,  Garrido R.,  Moya A.,  Garc\'ia~Hern\'andez
  A.,  Rod\'on J.,   Lares-Martiz M.,  2018, \mn@doi [A\&A (in press)]
  {10.1051/0004-6361/201732431}

\bibitem[\protect\citeauthoryear{Pelletier}{Pelletier}{1997}]{Pelletier97}
Pelletier J.~D.,  1997, Physical review letters, 78, 2672

\bibitem[\protect\citeauthoryear{Poretti et~al.,}{Poretti
  et~al.}{2009}]{Poretti09HD5084}
Poretti E.,  et~al., 2009, Astronomy \& Astrophysics, 506, 85

\bibitem[\protect\citeauthoryear{Provenzale, Smith, Vio  \& Murante}{Provenzale
  et~al.}{1992}]{Provenzale92}
Provenzale A.,  Smith L.,  Vio R.,   Murante G.,  1992, Physica D: Nonlinear
  Phenomena, 58, 31

\bibitem[\protect\citeauthoryear{Provenzale, Villone, Babiano  \&
  Vio}{Provenzale et~al.}{1993}]{Provenzale93}
Provenzale A.,  Villone B.,  Babiano A.,   Vio R.,  1993, International Journal
  of Bifurcation and Chaos, 03, 729

\bibitem[\protect\citeauthoryear{{Roudier} \& {Muller}}{{Roudier} \&
  {Muller}}{1986}]{SolGran86}
{Roudier} T.,  {Muller} R.,  1986, \mn@doi [\solphys] {10.1007/BF00155337},
  \href {http://adsabs.harvard.edu/abs/1986SoPh..107...11R} {107, 11}

\bibitem[\protect\citeauthoryear{Stanley \& Wong}{Stanley \&
  Wong}{1972}]{Stanley72}
Stanley H.~E.,  Wong V.~K.,  1972, American Journal of Physics, 40, 927

\bibitem[\protect\citeauthoryear{Tarboton, Bras  \& Rodriguez-Iturbe}{Tarboton
  et~al.}{1988}]{Tarboton88}
Tarboton D.~G.,  Bras R.~L.,   Rodriguez-Iturbe I.,  1988, Water Resources
  Research, 24, 1317

\bibitem[\protect\citeauthoryear{Torrence \& Compo}{Torrence \&
  Compo}{1998}]{TorrenceWave}
Torrence C.,  Compo G.~P.,  1998, Bulletin of the American Meteorological
  Society, 79, 61

\bibitem[\protect\citeauthoryear{Uytterhoeven et~al.,}{Uytterhoeven
  et~al.}{2011}]{Uytterhoeven2011}
Uytterhoeven K.,  et~al., 2011, \mn@doi [Astronomy and Astrophysics]
  {10.1051/0004-6361/201117368}, 534, 125

\bibitem[\protect\citeauthoryear{{Yamamoto} \& {Hughson}}{{Yamamoto} \&
  {Hughson}}{1993}]{Yama}
{Yamamoto} Y.,  {Hughson} R.~L.,  1993, \mn@doi [Physica D Nonlinear Phenomena]
  {10.1016/0167-2789(93)90083-D}, \href
  {http://adsabs.harvard.edu/abs/1993PhyD...68..250Y} {68, 250}

\bibitem[\protect\citeauthoryear{Zhou, Feng, Wu, Li  \& Liu}{Zhou
  et~al.}{2014}]{zhou14}
Zhou S.,  Feng Y.,  Wu W.-Y.,  Li Y.,   Liu J.,  2014, Research in Astronomy
  and Astrophysics, 14, 104

\makeatother
\end{thebibliography}


\appendix

\section{Supplementary Material}

\subsection{Fractal analysis with simple theoretical models}\label{sec:Frac_Models}

\begin{figure}
	\centering
	\includegraphics[width=9cm]{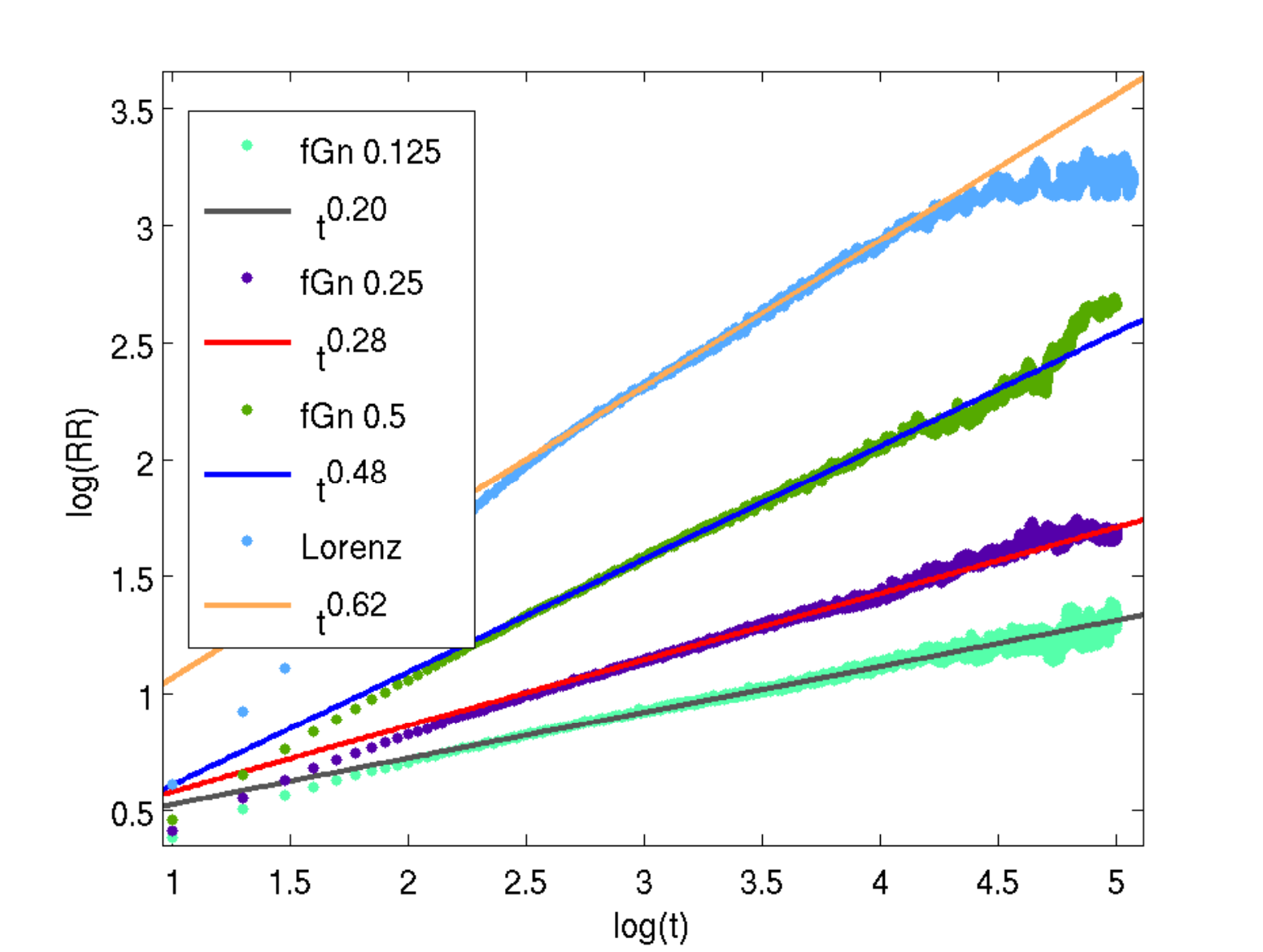}
	\caption{R/S analysis for different fGn time series and for the $x(t)$ variable of 				a Lorenz model.}
	\label{RR_fGn_Lorenz}
\end{figure}

The most simple model of self affine time series is the Gaussian white noise, i.e. a series $y_{i}=y(t_i)$ where each value is extracted from a Gaussian distribution. In this case the mean value $\left<y_i^2\right>$ , performed over all the possible realizations of $y_i$ at time $t_{i}$, is constant over time. From the self affine equation \ref{uno} $H=0$, moreover, since the series is totally uncorrelated its Fourier Spectra is flat, i.e. $\beta=0$ , see equation \ref{PS}. The time series of the displacement of a random walk, the so called Brownian motion, could be obtained by integration of a Gaussian white noise, and has a mean square displacement $\left<y(t)^2\right>\propto t$, thus $H=\frac{1}{2}$ and, since $\beta=2H+1$, $\beta=2$.  More in general it is possible to generate series with different values of $\beta$ by Fourier coefficient filtering and time integration of Gaussian white noise, obtaining stationary Fractional Gaussian Noises (fGn), with $\beta\in\left[-1,1\right]$ and non-stationary Fractional Brownian Motions (fBm), with $\beta\in\left[1,3\right]$ ($H\in\left[ 0,1\right]$) \citep{Turco}.  When $H\in\left[0,1\right]$, a self affine curve is also a geometrical fractal, and the box counting fractal dimension $D=2-H$; for instance the  fractal dimension of a Brownian motion is $D=\frac{3}{2}$, something between a line and a square.\\
When we study fGn with $\beta\in\left[-1,1\right]$,  $H\simeq0$ and there is no more linearity between $H$ and $\beta$. Thus, in order to characterize a fGn process one needs to transform it into fBm, by series integration, and then perform the R/S analysis, with exponent $\alpha\simeq \frac{\beta+1}{2}$.\\
On the other hand with R/S analysis it is possible to find the fractal regimes also in deterministic chaos, like for instance in Lorenz model, one of the first models of chaos describing the turbulent dynamic of convective cells \citep{Lorenz63}. In Fig.~\ref{RR_fGn_Lorenz} some examples of R/S analysis with $N\simeq 10^5$ long fGn and Lorenz model\footnote{  $\frac{dx}{dt}=\sigma(y - x)$,$\frac{dy}{dt}=x(\rho-z)-y$ and $\frac{dz}{dt}=xy-\beta z$, with parameters $\sigma=10$, $\beta=8/3$ and $\rho=28$} time series are shown.

\subsection{Multifractal analysis}\label{sec:MSS}
\begin{figure}
	\centering
	\includegraphics[width=9cm]{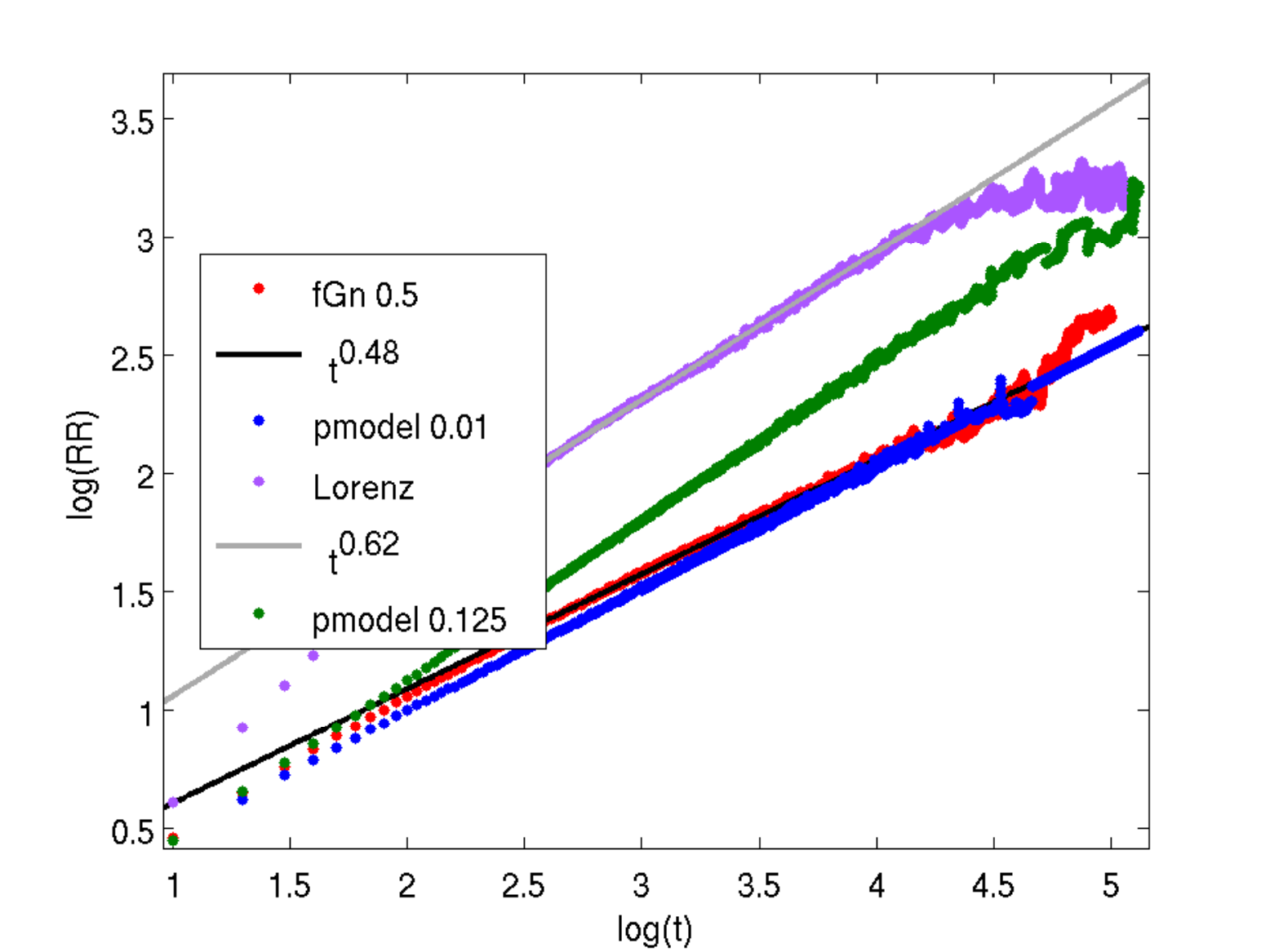}
	\caption{RR analysis for Gaussian noise, p model, and Lorenz model. 				$\alpha$ exponent for fGn and p model (p=0.01) and for Lorenz 				model and p model (p=0.125) are similar, although their 					Multifractal Singularity Spectra shows their different level of 	multifractality and complexity (see next figure)}
	\label{RR_pmodel}
\end{figure}

\begin{figure}
	\centering
	\includegraphics[width=9cm]{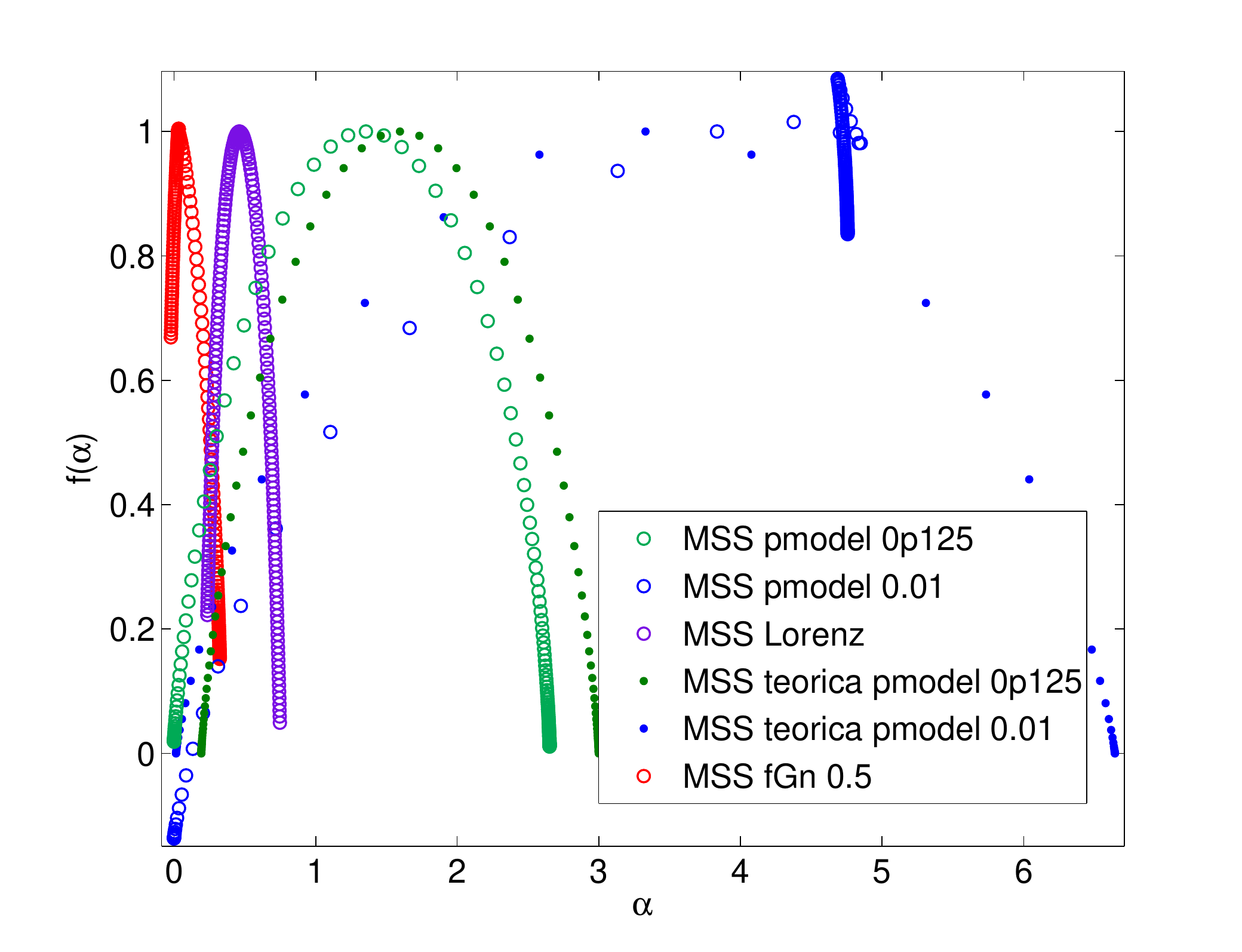}
	\caption{MSS analysis for Gaussian noise, p model, and Lorenz model. For all stochastic time series we use WTMM algorithm with $m=3$ order Morlet base, in a  range of $t\in[100,10000]$,  while $\kappa\in(0.15,0.25)$. For Lorentz time series $t\in[1000,3162]$ 	and $\kappa=1$. Theoretical MSS  for p model is described in \protect{\citep{Cheng14}}}
	\label{MSS_pmodel}
\end{figure}

\begin{figure}
	\centering
	\includegraphics[width=9cm]{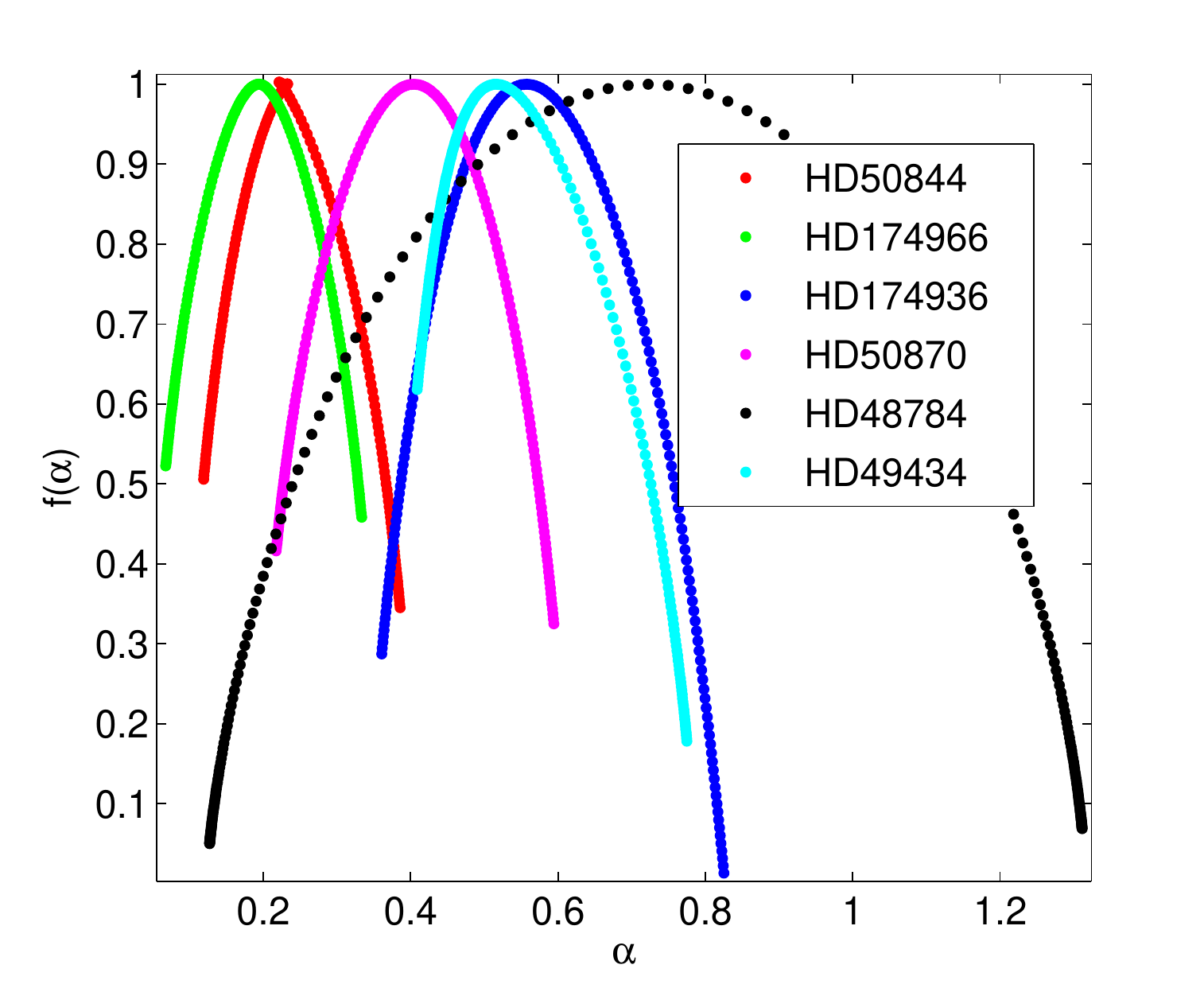}
	\caption{Multifractal singularity spectrum analysis for the 6 CoRoT $\delta$-Scuti stars. We employed the wavelet WTMM algorithm. Details on algorithm parameters are specified in table \ref{tab-technical}.}
	\label{MF_dScuti}
\end{figure}

\begin{figure*}
	\centering
	\includegraphics[width=7cm]{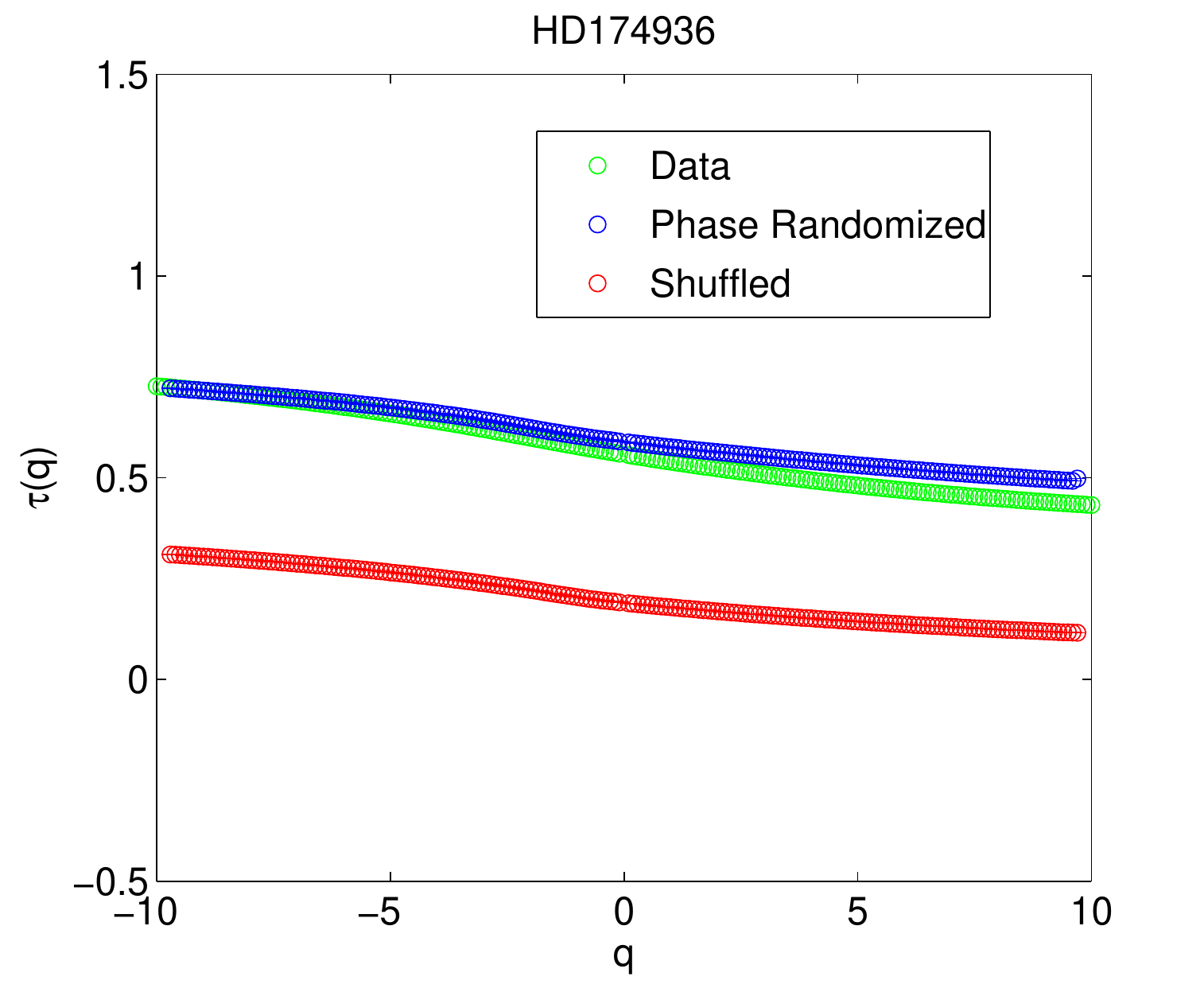}
	\includegraphics[width=7cm]{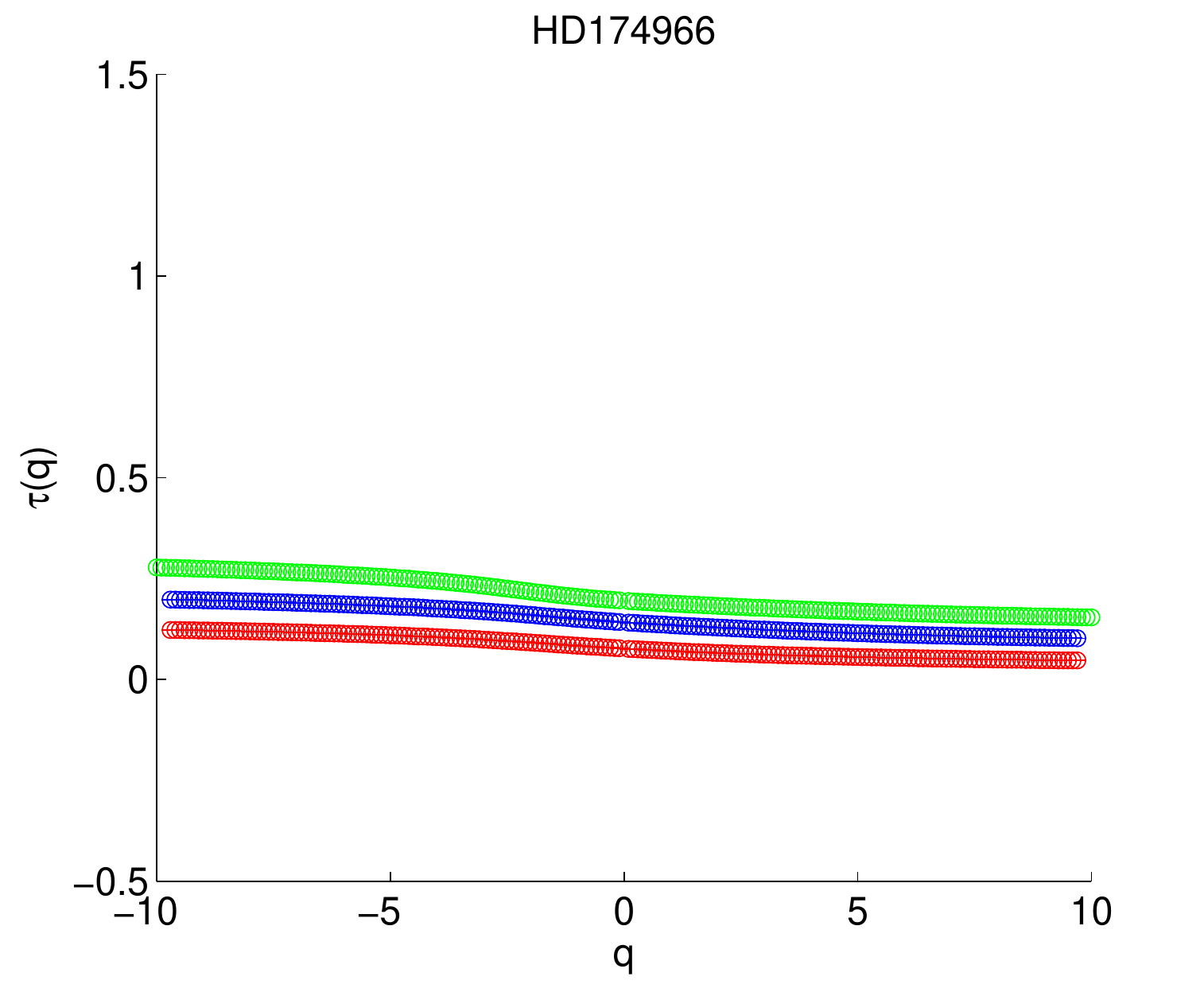}
	\includegraphics[width=7cm]{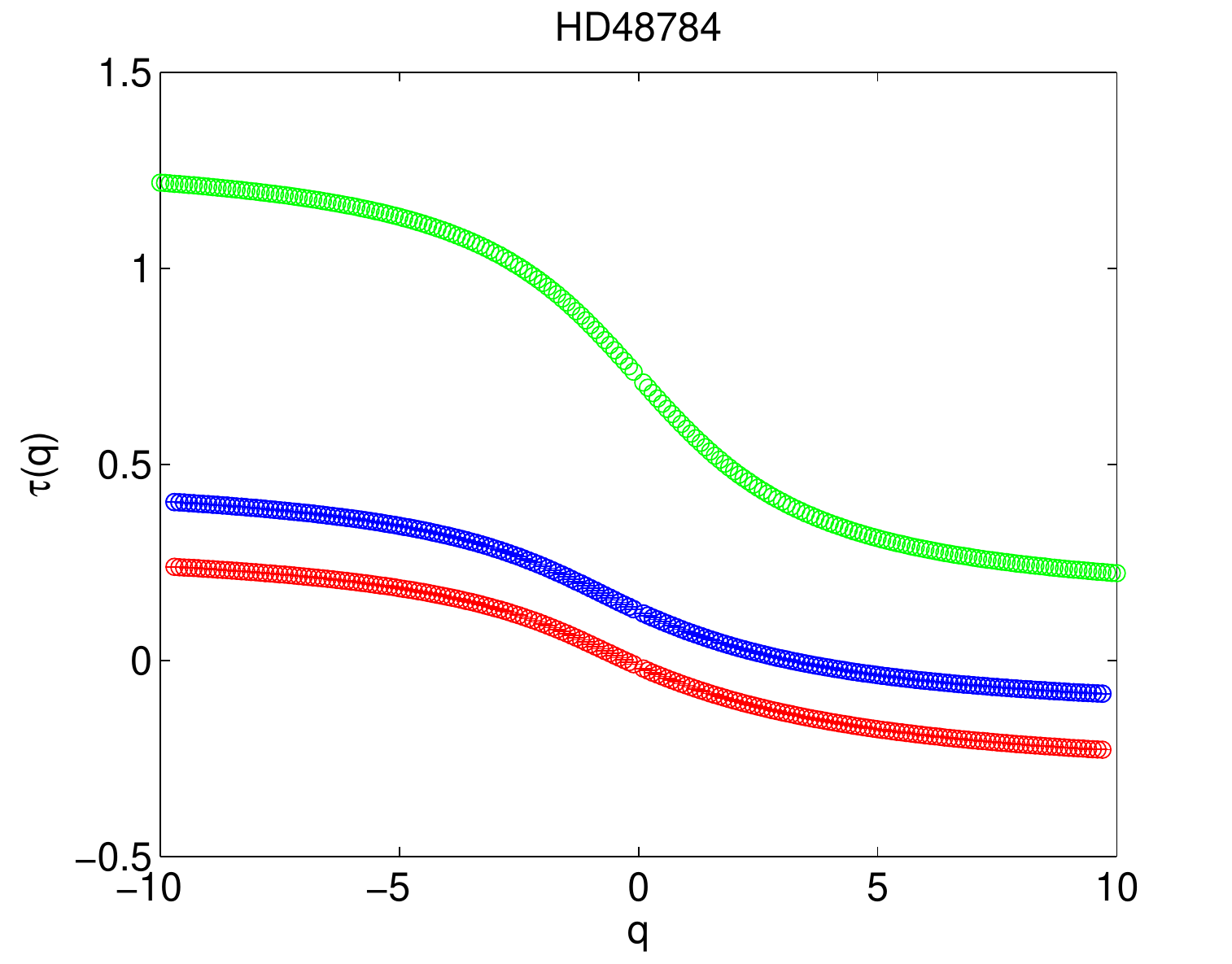}
	\includegraphics[width=7cm]{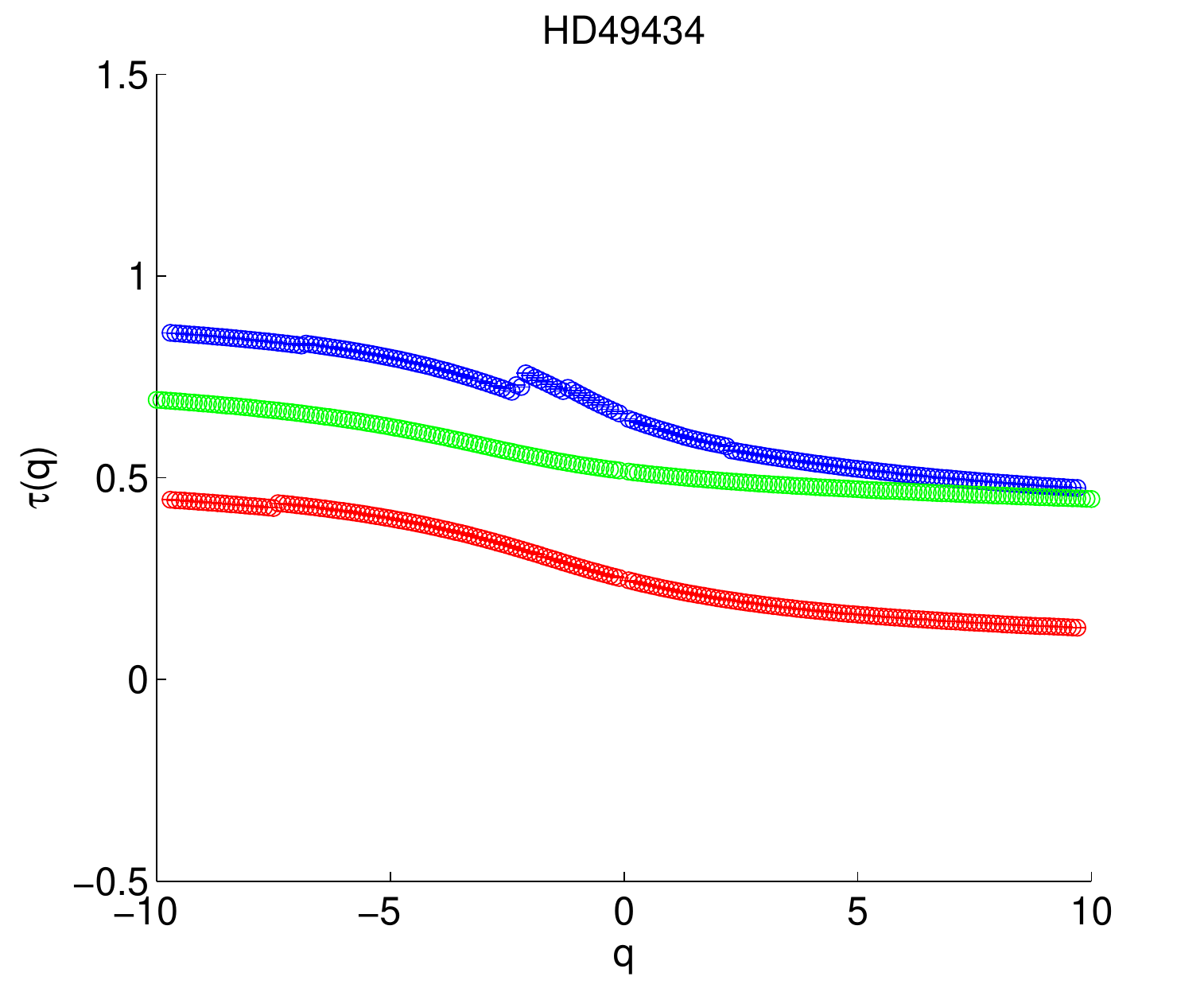}
	\includegraphics[width=7cm]{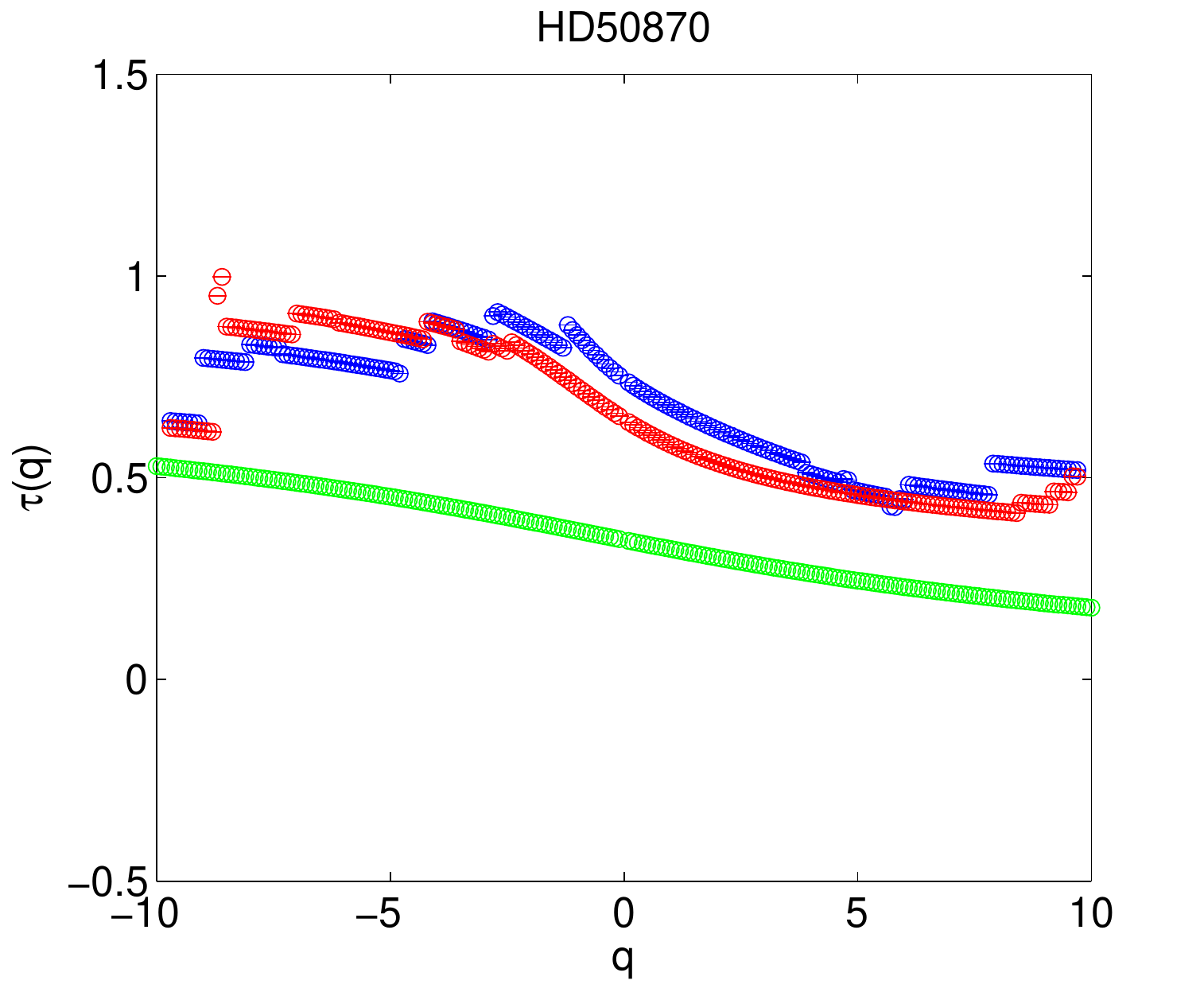}
	\caption{ Profiles $\hat{\tau}(q)$ for the original series (green curves), for Shuffled (red curves) and for Phase Randomized (blue curves) surrogate series.}\label{tau_vs_q}
\end{figure*}

Some time series do not exhibit a simple monofractal scaling behavior, which can be accounted for by a single $\alpha$ scaling exponent, and such different scaling behavior can be observed for many interwoven fractal subsets of the time series. Thus a multitude of scaling exponents, associated with different behaviors of small and large fluctuations, is required for a full description of the scaling behavior and a multifractal analysis must be applied \citep{Kantel08}.\\
Multifractal singularity spectrum (MSS) gives us the whole broad range of $\bar{\alpha}$ scaling exponents with their relative weights. We numerically compute MSS exploiting Wavelet Analysis.\\
Similarly to the the Fourier transform, the wavelet transform of a discrete time series $y(i)$ is a convolution sum:
\eqna{\label{intWave}
L_{\psi}(\tau,s)=\frac{1}{s}\sum_{i=1}^{N}y(i)\psi\left[ (i-\tau)/s \right],
}
where $\psi(t)$ is the mother wavelet, from which all the transform basis functions, i.e. the daughter wavelets $\psi\left[ (i-\tau)/s \right]$, are generated by shifting and stretching of the time axis. The wavelet coefficients $L_{\psi}(\tau,s)$ thus depend on both time position $\tau$ and scale s. Hence, the local frequency decomposition of the signal is described with a time resolution appropriate for the considered frequency $f=1/s$. Wavelet transform performs a sort of local, temporally windowed, Fourier transform. All wavelets $\psi(t)$ must have zero mean. They are often chosen to be orthogonal to polynomial trends, so that
the analysis method becomes insensitive to possible trends in the data. In our case we use the Morlet wavelet with complex $m$ order:
\eqna{\label{Morlet}
\psi_{0}(\eta)=\pi^{\frac{1}{4}}e^{im\eta}e^{-\eta^{2}/2}.
}
In order to compute the MSS we implemented the wavelet transform modulus maxima (WTMM) method, a well-known method to investigate the multifractal scaling properties of fractal and self affine series in the presence of non-stationarities \citep{Kantel08,TorrenceWave}. 
Note that in this case the series $y(i)$ are analyzed directly instead of the cumulative profile $Y(n)$ defined in fluctuation analysis. 
In this method, instead of averaging over all wavelet coefficients, one averages, within the modulo-maxima method only the local maxima of $|L_{\psi}(\tau, s)|$. First, one determines for a given scale $s$, the positions $\tau_j$ of the local maxima of $|L_{\psi}(\tau, s)|$ as a function of $\tau$, so that $|L_{\psi}(\tau_j-1,s)|<|L_{\psi}(\tau_j, s)|\geq|L_{\psi}(\tau_j+1,s)|$ for $j=1,\ldots,j_{max}$. Then one sums 
up the qth power of the maxima,
\eqna{\label{Zwave}
Z(q,s)=\sum_{j=1}^{j_{max}}|L_{\psi}(\tau_j, s)|^{q}
}
The reason for the maxima procedure is that the absolute wavelet coefficients $|L_{\psi}(\tau_j, s)|$ can become arbitrarily small. The analyzing wavelet $\psi(x)$ must always have positive values for some $x$ and negative values for other $x$, since it has to be orthogonal to possible constant trends. Hence there are always positive and negative terms in the sum of equation \ref{intWave}, and these terms might cancel. If that happens $|L_{\psi}(\tau_j, s)|$ can become close to zero. Since such small terms would spoil the calculation of negative moments in equation \ref{Zwave}, they have to be eliminated by the maxima procedure.\\
The values $|L_{\psi}(\tau_j, s)|$ might become smaller for increasing s since just more (positive and negative) terms are included in the sum of equation \ref{intWave}, and these might cancel even better. Thus, an additional supremum procedure has been introduced in the WTMM method in order to keep the dependence of $Z(q,s)$ on $s$ monotonous. If, for a given scale $s$, a maximum at a certain position $\tau_j$ happens to be smaller than a maximum at $\tau^{'}_{j}\approx\tau_{j}$ for a lower scale $s'<s$, then $|L_{\psi}(\tau_j, s)|$ is replaced by $|L_{\psi}(\tau^{'}_j, s)|$ in the sum \ref{Zwave}. In our criteria we compared the maximum at different scales $s'<s$ only if their temporal distance is
\eqna{\label{k_correction}
	|\tau_j-\tau^{'}_j|<\kappa \frac{N(y)}{N_{max}(s^{'})},
}
where $N(y)$ is the total length of the series, $N_{max}(s^{'})$ is the number of maxima at $s^{'}$ scale and $\kappa$ is a multiplicative factor.\\
Scaling behavior is observed for $Z(q,s)$  and scaling exponents $\hat{\tau}(q)$ can be defined that describe how $Z(q,s)$ scales with s,
\eqn{
Z(q, s)\sim s^{\hat{\tau}(q)}
}\label{Zexptau} 
The exponents $\hat{\tau}(q)$ characterize the multifractal properties of the series under investigation.\\
Multifractal singularity spectrum $f(\bar{\alpha})$ is related to $\hat{\tau}(q)$ via a Legendre transform,
\eqna{\label{legendre}
	\bar{\alpha}&=&\frac{d}{dq}\hat{\tau}(q),\\
	f(\bar{\alpha})&=&q\bar{\alpha}-\hat{\tau}(q)
}
Here $\bar{\alpha}$ is the singularity strength or local Hurst exponent, while $f(\bar{\alpha})$ denotes the dimension of the subset of the series that is characterized by $\bar{\alpha}$. Such multifractal approach can be considered as a generalized version of the fluctuation analysis method, that make use of the second order fluctuation to find the standard (mono) fractal self-affine exponent in eq.\ref{uno}, i.e. $\alpha=\frac{1+\tau(2)}{2}$. A measure of the degree of multifractality considers the range of variation of $\alpha$ local Hurst exponents involved in the time series, i.e. $\Delta \alpha$. \\
The most typical example of model showing a clear multifractality behaviour is the p-model, base on the binomial cascade \citep{Meneveau87, drozdz,Cheng14}. In figures \ref{RR_pmodel} and \ref{MSS_pmodel} there is a clear evidence that time series apparently similar in their fractal behaviour under R/S analysis, like fGn ($\alpha=0.5$) and p model (p=0.01), have different degrees of multifractality, so that p model with $\Delta \alpha\sim 6$ is a developed multifractal, while fGn, with $\Delta \alpha\sim 0.5$, is most likely a mono fractal.
The MSS for the 6 \dss\ reported in table \ref{tab:Stars} gives us some evidences of a mutifractal behaviour, in order of decreasing $\Delta\alpha$ for HD48784,  HD49434, HD174936 and HD 50870, while HD174966, HD50844 seems to be essentially monofractals, see figure \ref{MF_dScuti} and table \ref{tab-technical} for technical details. We are confident that MSS analysis, coupled with $ \fractal$, R/S and power spectrum analysis could allow us to better understand modes deterministic/stochastic/chaotic dynamic and their correspondent excitation mechanisms, depending in some cases on the convective region of the star. However MSS implementation on real data resulted in our work to be very unstable and with a strong dependence on algorithms and their parameters, then further numerical and theoretical work is needed. \\
\begin{table}
	\centering
	\caption{Temporal ranges, measured in days, considered for the RR fit and the calculation of MSS with the wavelet WTMM method. For the 	WTMM method the order of Morlet mother wave and the $\kappa$ parameters involved in the peaks correction window size, equation \ref{k_correction} are specified.}
    \label{tab-technical}
	\begin{tabular}[!h]{lccc}
\hline
ID & RR fit / MSS WTMM range & Morlet order & $\kappa$\\
\hline
HD174936 & $0.117-1.171$  & $6$ & $1.5$ \\ 
HD174966 & $0.370-3.704$  & $6$ & $5.0$\\
HD48784  & $1.171-11.712$ & $3$  &none \\
HD49434  & $3.704-37.037$ & $3$ & $0.65$ \\
HD50844  & $0.370-11.712$ & $6$ & $1.0$ \\
HD50870  & $0.370-6.586$  & $6$ & $2.5$ \\
\hline
	\end{tabular}
\end{table}
%
%
In \cite{NewSunsIII,NewSunsIV} the authors studied the MSS of solar-like and M dwarfs light curves dataset employing shuffling and phase randomized surrogates procedures, in order to distinguish different sources of multifractality in the time series \citep{Provenzale93}. In fact two general types of multifractality in time series can be distinguished:\\ 
(i) Multifractality due to a broad probability distribution (density function) for the values of the time series, e. g. a Levy or power law distribution. In this case the multifractality cannot be removed by shuffling the series. For this case, the non-Gaussian effects can be weakened by creating phase-randomized surrogates. In this context, the procedure preserves the amplitudes of the Fourier transform and the linear properties of the original series but randomizes the Fourier phases while eliminating nonlinear effects.\\ 
(ii) Multifractality due to different long-term correlations of the small and large fluctuations. In this case the probability density function of the values can be a regular distribution with finite moments, e.g., a Gaussian distribution. The corresponding shuffled series will exhibit non-multifractal scaling, since all long-range correlations are destroyed by the shuffling procedure.\\
To separate the contributions to multifractality of the above described two different sources, it is a very useful technique to study the profiles $\hat{\tau}(q)$ in formula \ref{Zexptau} for the original series, for Shuffled and for Phase Randomized surrogate series. Amplitude $\hat{\tau}(q_{min})-\hat{\tau}(q_{Max})$, similarly to $\Delta\tilde{\alpha}$, gives a quantitative estimation of multifractality, while the difference in the profiles of  original series $\hat{\tau}(q)$ with shuffled and phase randomized ones gives us the contribution of sources (i)  and (ii) to multifractality, see figure \ref{tau_vs_q}. By way of example, it is possible to infer that in HD174936 the contribution of nonlinear dynamics and broadband distribution to multifractality are negligible, since blue and red profiles are almost identical, in HD49434 only the short fluctuations, detected by $q<0$ profiles have some nonlinear dynamic involved, while in the rest of curves contribution from long term correlations and nonlinear dynamics seems to be almost equally involved. We believe that a joint analysis of MSS, $\hat{\tau}(q)$ profiles and CGSA $\fractal$ could help to understand and separate quantitatively deterministic non-chaotic behaviour, nonlinear chaotic and stochastic dynamics. In order to obtain this goal we are working in progress on a deeper algorithm performance and data analysis. 

\subsection{Test on CGSA Algorithm sensitivity}\label{sec:CGSA_test}
In table \ref{tab:CGSA_fGn} we report the result of some test of the sensitivity of our CGSA algorithm, applied to a series made by the sum fo a fGn (with standard deviation $\sigma_{fGn}=1$) with an harmonic signal $A\sin(ft)$.  For the analysis of self-affine signals, CGSA algorithm is more stable with the introduction of an harmonic function, even with a small amplitude $A<<\sigma_{fGn}$,  i.e. the standard deviation of $\fractal$ becomes smaller compared with the pure fGn. This is a good point in the interpretation of our analysis, because in an actual light curve we will always have a mix of harmonic modes with self affine background. 
\begin{table}
    \centering
	\caption{Mean and standard deviation from the distributions obtained by 			$330$ runs of $L=36409$ long time series made by the sum of a 				fractional Gaussian noise, with Hurst exponent $\alpha$, with a 			harmonic function of amplitude $A$ and frequency $f=10/L$. }
    \label{tab:CGSA_fGn}
	\begin{tabular}[!h]{cccc}
\hline
$\alpha$ & A & $\mu(\%Frac)$ & $\sigma(\%Frac)$  \\
\hline
 $0.125$ & $0$ & $1.05$ & $0.15$ \\
 $0.125$ & $0.1$ & $1.01$ & $0.01$ \\
 $0.125$ & $1$ &  $0.696$ &  $0.007$ \\
 $0.125$ & $10$ & $0.051$ &  $0.001$ \\
  $0.5$   & $0$ & $1.07$ & $0.33$ \\
  $0.5$ & $0.1$ &  $1.11$ & $0.01$ \\
  $0.5$ & $1$ & $0.77$ & $0.01$ \\
  $0.5$   & $10$ &  $0.055$ & $0.002$ \\
  $0.875$ & $0$ & $1.$ & $0.3$ \\
  $0.875$ & $0.1$ & $1.08$ & $0.03$ \\
  $0.875$ & $1$ & $0.83$ & $0.06$ \\
 $0.875$ & $10$ & $0.082$ & $0.009$ \\
\hline
	\end{tabular}
\end{table}
%

\bsp	
\label{lastpage}
\end{document}